\documentclass[11pt]{article}
\usepackage{longtable}
\usepackage{ae,aecompl,pslatex}
\usepackage[dvips]{graphicx}
\usepackage{enumerate,color}
\usepackage{fullpage}
\usepackage{lineno}
\usepackage[table]{xcolor}
\definecolor{lightgray}{gray}{0.9}
\usepackage{float}

\usepackage{placeins,hyperref}
\usepackage{psfrag,epsfig,graphicx,color,array,dcolumn,tabularx,delarray,longtable,indentfirst,amsmath,amsfonts,amssymb,amsthm,eucal,bbm}
\theoremstyle{bkaexa} 
\usepackage{caption}
\usepackage{setspace}
\theoremstyle{bkaexa} 

\theoremstyle{bkathm} 

\theoremstyle{bkathm} 
\newtheorem{Thm}{Theorem}
\theoremstyle{bkathm} 

\theoremstyle{bkathm} 
\newtheorem{Lem}{Lemma}
\theoremstyle{definition}

\begin{document}
\setstretch{1.5}
\title{On the extensions of the Chatterjee-Spearman test}
\author{\normalsize Qingyang Zhang\\
\normalsize Department of Mathematical Sciences, University of Arkansas, AR 72701\\
\normalsize Email: qz008@uark.edu
}
\date{}
\maketitle

\begin{abstract}
Chatterjee (2021) introduced a novel independence test that is rank-based, asymptotically normal and consistent against all alternatives. One limitation of Chatterjee's test is its low statistical power for detecting monotonic relationships. To address this limitation, in our previous work (Zhang, 2024, \textit{Commun. Stat. - Theory Methods}), we proposed to combine Chatterjee's and Spearman's correlations into a max-type statistic and established the asymptotic joint normality. This work examines three key extensions of the combined test. First, motivated by its original asymmetric form, we extend the Chatterjee-Spearman test to a symmetric version, and derive the asymptotic null distribution of the symmetrized statistic. Second, we investigate the relationships between Chatterjee's correlation and other popular rank correlations, including Kendall's tau and quadrant correlation. We demonstrate that, under independence, Chatterjee's correlation and any of these rank correlations are asymptotically joint normal and independent. Simulation studies demonstrate that the Chatterjee-Kendall test has better power than the Chatterjee-Spearman test. Finally, we explore two possible extensions to the multivariate case. These extensions expand the applicability of the rank-based combined tests to a broader range of scenarios.
\end{abstract}


\section{Introduction}
Let $X$ and $Y$ be two continuous random variables, and $(X_{i},~ Y_{i})_{i=1,...,n}$ be $n$ independent and identically distributed ($i.i.d.$) samples of $(X, Y)$. In this work, we are interested in the following classical independence test, formulated as 
\begin{align*}
H_{0}&: X\perp Y,\\
H_{a}&: X \not\perp Y.
\end{align*}

The problem of testing independence has been examined from multiple perspectives, leading to the development of numerous testing methods. Rank-based methods, in particular, have gained increasing attention due to their nice properties such as distribution-freeness and B-robustness \cite{crouxdehon}. Spearman's and Kendall's rank correlations are widely used for measuring and detecting monotonic relationships between variables. However, they are not suitable for analyzing non-monotonic associations and may have low statistical power in such cases. To address this limitation, several novel rank tests have been proposed including Hoeffding's $D$ \cite{Hoeffding}, Bergsma-Dassios' $\tau^{*}$ \cite{BD}, and Blum-Kiefer-Rosenblatt's $R$ \cite{BKR}. These tests are consistent against all alternatives, meaning that they can detect both monotonic and non-monotonic associations. Recently, Chatterjee (2021) introduced a new correlation test that is also rank-based and consistent, but unlike the aforementioned tests, Chatterjee's test has a simple asymptotic theory which enables analytical calculation of p-value \cite{chatterjee}. Due to its nice properties, Chatterjee's test has been extensively studied including its asymptotic behavior, local power analysis, extensions and applications \cite{Zhang24, ShiHan21, LinHan23, auddy, azadkia.chatterjee, LinHan22, Zhang23, CaoBickel, Deb.etal, Huang, ShiHan24, HanHuang, chatterjeenet}. For a recent survey about this important method, see \cite{chatterjee.survey}.

We begin by the definition of Chatterjee's correlation. Assuming that $X_{i}$'s and $Y_{i}$'s have no ties, the $i.i.d.$ samples can be uniquely arranged as $(X_{(1)},~ Y_{(1)}),~ ...,~ (X_{(n)},~ Y_{(n)})$, such that $X_{(1)}<~\cdots<~X_{(n)}$. Here $Y_{(1)},~ ...,~ Y_{(n)}$ denote the concomitants. Let $R_{i}$ be the rank of $Y_{(i)}$, i.e., $R_{i} = \sum_{k=1}^{n}\mathbbm{1}\{Y_{(k)}\leq Y_{(i)}\}$, Chatterjee's correlation is defined as
\begin{equation}
\xi_{n}(X,~Y) = 1-\frac{3\sum_{i=1}^{n-1}|R_{i+1}-R_{i}|}{n^2-1}.
\end{equation}
Chatterjee (2021) showed that, provided that $Y$ is not a constant almost surely, $\xi_{n}(X, Y)$ converges almost surely to the following population quantity
\begin{equation}
\xi(X,~Y) = \frac{\int V(E(\mathbbm{1}\{Y\geq t|X\}))dF_{Y}(t)}{\int V(\mathbbm{1}\{Y\geq t\})dF_{Y}(t)},
\end{equation}
which is known as the Dette-Siburg-Stoimenov (DSS) measure \cite{dss}. Notably, the DSS measure is 0 if and only if $X$ and $Y$ are independent, and 1 if $Y$ is a measurable function of $X$. Both $\xi_{n}(X,~Y)$ and $\xi(X,~Y)$ are asymmetric, and a symmetrized version of $\xi_{n}(X,~Y)$ is studied in \cite{Zhang23}. 

Under independence, i.e., $\xi(X,~Y) = 0$, one can show $E[\xi_{n}(X,~Y)] = 0$. Zhang (2023) derived the finite-sample variance 
\begin{equation*}
V[\sqrt{n}\xi_{n}(X,~Y)] = \frac{n(n-2)(4n-7)}{10(n+1)(n-1)^2}.
\end{equation*}
Furthermore, as $n\rightarrow\infty$, $\sqrt{n}\xi_{n}(X, Y)$ converges to a normal distribution with mean 0 and variance $2/5$ \cite{chatterjee}. Lin \& Han (2022) established the central limit theorem of $\xi_{n}(X,~Y)$ under arbitrary dependence structures, as long as $Y$ is not a measurable function of $X$. 

While $\xi_{n}(X,~Y)$ is sensitive to non-monotonic associations, particularly those with oscillating patterns like sinusoids, it exhibits substantially lower power for monotonic associations compared to other rank tests such as $D$, $R$ and $\tau^{*}$. This limitation of Chatterjee's test could be a practical concern, prompting recent efforts to improve its power. For instance, Lin \& Han (2023) constructed a new test statistic by incorporating multiple right nearest neighbors. In our previous work \cite{Zhang24}, we proposed to combine Chatterjee's and Spearman's correlations, where the former is sensitive to non-monotonic associations while the latter is powerful in detecting monotonic associations. Moreover, we established the asymptotic independence and joint normality of the two correlations being combined. The Spearman correlation, denoted by $S_{n}(X, Y)$, is defined as
\begin{equation*}
S_{n}(X, Y) = 1 - \frac{6\sum_{i=1}^{n}(i-R_{i})^2}{n(n^2-1)}.
\end{equation*}
Under independence, we have
$$\sqrt{n}\begin{bmatrix} 
S_{n}(X, Y)\\
\xi_{n}(X, Y) 
\end{bmatrix}
\xrightarrow{d} N\left [
\begin{pmatrix}
0\\
0
\end{pmatrix}, 
\begin{pmatrix}
1 & 0 \\
0 & 2/5 
\end{pmatrix} \right ],$$ as $n\rightarrow\infty$. This motivated us to define the following max-type statistic that combines the strengths of both correlations
\begin{equation*}
I_{n}(X,~Y) = \max\{|S_{n}(X,~Y)|, \sqrt{5/2}\xi_{n}(X,~Y)\}.
\end{equation*}

While our previous simulation studies demonstrated the effectiveness of $I_{n}(X,~Y)$ under various correlation patterns, several key aspects remain unexplored. First, same as $\xi_{n}(X,~Y)$, the combined measure $I_{n}(X,~Y)$ is asymmetric. Misspecifying the order of $(X,~Y)$ may lead to reduced power, particularly for non-monotonic relationships. This is especially critical in applications like gene co-expression analysis, where symmetric tests are generally preferred. Second, Zhang (2024) focused on combining Chatterjee's correlation with Spearman's correlation. How does it perform with other rank correlations such as Kendall's tau or quadrant correlation? Third, the current Chatterjee-Spearman test is limited to random scalars. Can we extend this framework to handle random vectors? 

This paper addresses these limitations by proposing several extensions. Section 2 introduces the new tests derived from our framework, including a symmetric version of the Chatterjee-Spearman test, the Chatterjee-Kendall test, and the Chatterjee-quadrant test. Their multivariate counterparts are also explored. Section 3 evaluates the performance of these new tests under various scenarios using simulation studies. Section 4 applies different tests to two real-world datasets to demonstrate their practical utility. Section 5 discusses and concludes the paper with some future perspectives.

\section{Method}
\subsection{Extension to the symmetric case}
The asymmetric test based on $I_{n}(X,~Y)$ requires specifying a response variable $Y$ and an independent variable $X$. Switching $X$ and $Y$ may lead to different test results. We give an example here. Let $X$ be a uniform random variable on $[-1,~1]$ and $Y = |X| + 0.1Z$, where $X\perp Z$ and $Z\sim N(0,~1)$. For $n = 100$, we have $I_{n}(X,~Y) \approx 1.072$ ($\mbox{p-value} \approx 0$) while $I_{n}(Y,~X) \approx 0.061$ ($\mbox{p-value} \approx 0.667$). 

To construct a symmetric test, we need to derive the asymptotic joint distribution of $S_{n}(X,~Y)$, $\xi_{n}(X,~ Y)$ and $\xi_{n}(Y,~ X)$ under independence. The following two lemmas provide the asymptotic covariances, indicating that the three components are asymptotically uncorrelated.  

\begin{Lem}[Lemma 1 in \cite{Zhang24}]\label{lem1}
If $X$ and $Y$ are independent, we have
\begin{equation*}
\mbox{Cov}\left[ S_{n}(X,~ Y), \xi_{n}(X,~ Y) \right ] = 0,
\end{equation*}
for any $n\geq 2$. 
\end{Lem}

\begin{Lem}[Corollary 1 in \cite{Zhang23}]\label{lem2}
If $X$ and $Y$ are independent, we have
\begin{equation*}
\mbox{Cov}\left[ \sqrt{n}\xi_{n}(X,~ Y),~ \sqrt{n}\xi_{n}(Y,~ X) \right ] = O(1/n).
\end{equation*}
\end{Lem}

Furthermore, using Cram\'{e}r-Wold device, we show that $S_{n}(X,~Y)$, $\xi_{n}(X,~ Y)$ and $\xi_{n}(Y,~ X)$ are asymptotically joint normal (the detailed proof is lengthy and we give it in Appendix A.1). Together with Lemmas \ref{lem1} and \ref{lem2}, we have the following theorem 
\begin{Thm} \label{thm1}
If $X$ and $Y$ are independent, we have
$$\sqrt{n}\begin{bmatrix}
S_{n}(X,~ Y)\\
\xi_{n}(X, ~Y) \\
\xi_{n}(Y,~ X)
\end{bmatrix}
\xrightarrow{d} N\left [
\begin{pmatrix}
0\\
0\\
0
\end{pmatrix}, 
\begin{pmatrix}
1 & 0 & 0 \\
0 & 2/5 & 0 \\
0 & 0 & 2/5 
\end{pmatrix} \right ],$$ as $n\rightarrow\infty$.
\end{Thm}

Our Theorem \ref{thm1} generalizes the main results in \cite{Zhang24} and \cite{Zhang23}. Similar to \cite{Zhang24}, we consider the following max-type statistic
\begin{equation*}
\widetilde{I}^{S}_{n}(X, Y) = \max\{|S_{n}(X, Y)|, \sqrt{5/2}\xi_{n}(X, Y), \sqrt{5/2}\xi_{n}(Y, X)\}.
\end{equation*}
It is easy to see that $\widetilde{I}^{S}_{n}(X, Y)$ is both nonnegative and symmetric, i.e., $\widetilde{I}^{S}_{n}(X, Y) = \widetilde{I}^{S}_{n}(Y, X)$. Moreover, using Theorem \ref{thm1}, we can calculate its asymptotic p-value:
\begin{align*}
P(\sqrt{n}\widetilde{I}^{S}_{n}(X, Y) > z) & \approx 1 - \Phi^{2}(z) \left[1-2\Phi(-z)\right] \\
& = 1+\Phi^{2}(z)-2\Phi^{3}(z),
\end{align*}
where $z\geq 0$ and $\Phi(\cdot)$ represents the cumulative distribution function ($c.d.f.$) of the standard normal distribution.

\subsection{Extension to other rank correlations}
This section explores the relationships between Chatterjee's correlation and some other rank correlations suitable for constructing new tests. We focus on Kendall's correlation and the quadrant correlation (also called quadrant count ratio), because, like Spearman's correlation, they capture monotonic relationships between variables. These two rank correlations are defined as follows
\begin{align*}
\tau_{n}(X,~ Y) & = \frac{2}{n(n-1)}\sum_{i<j}\mbox{sgn}\left[(X_{i} - X_{j})(Y_{i} - Y_{j})\right],\\
Q_{n}(X,~ Y) & = \frac{1}{n}\sum_{i=1}^{n}\mbox{sgn}\left[(X_{i} - \mbox{med}_{n}(X))(Y_{i} - \mbox{med}_{n}(Y))\right],
\end{align*}
where $\mbox{med}_{n}(X)$ represents the sample median of $(X_{1},~...,~X_{n})$. Their population counterparts are 
\begin{align*}
\tau(X,~ Y) & =  E\left\{ \mbox{sgn}\left[(X_{1} - X_{2})(Y_{1} - Y_{2})\right] \right\}, \\
Q(X,~ Y) & =  E\left\{ \mbox{sgn}\left[(X -  \mbox{med}(X))(Y -  \mbox{med}(Y))\right] \right\},
\end{align*}
where $(X_{1},~Y_{1})$ and $(X_{2},~Y_{2})$ represent two independent copies of $(X,~Y)$, and $\mbox{med}(X)$ is the median of $X$.

Under independence and a sample size of $n$, it is well-known that $E[\tau_{n}(X,~ Y)] = E[Q_{n}(X,~ Y)] = 0$. The variance of $\sqrt{n}\tau_{n}(X,~ Y)$ is provided by \cite{Han17}
\begin{equation*}
V[\sqrt{n}\tau_{n}(X,~ Y)] = \frac{2(2n+5)}{9(n-1)}.
\end{equation*}

The finite-sample variance of the quadrant correlation, to the best of our knowledge, has not been reported in the literature, therefore we derive the formula in Lemma \ref{lem3}. Interestingly, due to different sample median definitions for odd and even sample sizes, the variances exhibit slight discrepancies, although both converge to 1 as $n\rightarrow\infty$. The proof for Lemma \ref{lem3} is provided in Appendix A.2.
\begin{Lem}\label{lem3} 
If $X$ and $Y$ are independent, we have
\begin{equation*}
V[\sqrt{n}Q_{n}(X, Y)] = \begin{cases}
			(n-1)/n, & n\text{ is odd}\\
            n/(n-1), & n\text{ is even},
		 \end{cases}
\end{equation*}
for any $n\geq 2$.
\end{Lem}

To investigate their asymptotic joint distribution, we first establish in Lemma \ref{lem4} (proven in Appendix A.3) that $\xi_{n}(X,~Y)$, $\tau_{n}(X,~Y)$ and $Q_{n}(X,~Y)$ are uncorrelated for any sample size.
\begin{Lem}\label{lem4}
If $X$ and $Y$ are independent, we have
\begin{align*}
\mbox{Cov}\left[ \tau_{n}(X, Y),~ \xi_{n}(X, Y) \right ] & = 0 \\
\mbox{Cov}\left[ Q_{n}(X, Y),~ \xi_{n}(X, Y) \right ] & = 0
\end{align*}
for any $n\geq 2$. 
\end{Lem}

While Lemma \ref{lem4} demonstrates the uncorrelatedness between $\xi_{n}(X,~Y)$, $\tau_{n}(X,~Y)$ and $Q_{n}(X,~Y)$, it is important to note that these statistics are generally dependent. For instance, under independence and $n=3$, we have $\mbox{Cov}\left[ |\tau_{n}(X, Y)|,~ \xi_{n}(X, Y) \right ] = 1/18$, indicating that $\xi_{n}(X,~Y)$ and $\tau_{n}(X,~Y)$ are uncorrelated but dependent (see derivations in Appendix A.4). To establish the asymptotic independence, we also need to prove their asymptotic joint normality (similar to Theorem \ref{thm1}). Theorem \ref{thm2} below gives the asymptotic joint distributions for both $\{\sqrt{n}\tau_{n}(X, Y),~\sqrt{n}\xi_{n}(X, Y),~\sqrt{n}\xi_{n}(Y, X)\}$ and $\{\sqrt{n}Q_{n}(X, Y),~\sqrt{n}\xi_{n}(X, Y),~\sqrt{n}\xi_{n}(Y, X)\}$ (proof in Appendix A.5). The proofs of Theorem \ref{thm1} and \ref{thm2} are both based on Cr\'amer-Wold device and Chatterjee's central limit theorem \cite{chatterjee.clt}. The only difference lies in the asymptotic representations of Spearman's correlation and quadrant correlation (by \cite{Han17}, Kendall's correlation is asymptotically equivalent to Spearman's correlation, up to a constant, as we discussed in Appendix A.2). This difference leads to different expressions and bounds of the asymptotic variance in Chatterjee's central limit theorem.
\begin{Thm} \label{thm2}
If $X$ and $Y$ are independent, we have
$$\sqrt{n}\begin{bmatrix}
\tau_{n}(X, Y)\\
\xi_{n}(X, Y) \\
\xi_{n}(Y, X)
\end{bmatrix}
\xrightarrow{d} N\left [
\begin{pmatrix}
0\\
0\\
0
\end{pmatrix}, 
\begin{pmatrix}
4/9 & 0 & 0 \\
0 & 2/5 & 0 \\
0 & 0 & 2/5
\end{pmatrix} \right ],$$ and 
$$\sqrt{n}\begin{bmatrix}
Q_{n}(X, Y)\\
\xi_{n}(X, Y) \\
\xi_{n}(Y, X)
\end{bmatrix}
\xrightarrow{d} N\left [
\begin{pmatrix}
0\\
0\\
0
\end{pmatrix}, 
\begin{pmatrix}
1 & 0 & 0 \\
0 & 2/5 & 0 \\
0 & 0 & 2/5
\end{pmatrix} \right ],$$
as $n\rightarrow\infty$.
\end{Thm}

Figures 1 and 2 visually demonstrate the joint distributions of $\sqrt{n}\tau_{n}(X, Y)$ and $\sqrt{n}\xi_{n}(X, Y)$ (and $\sqrt{n}Q_{n}(X, Y)$ and $\sqrt{n}\xi_{n}(X, Y)$) under the null hypothesis and sample sizes 30, 100, 300 and 500, revealing a decreasing dependence between two statistics as the sample size increases. For larger sample sizes, such as $n=500$, the two statistics behave like independent variables.

\begin{center}
[Figure 1 about here]
\end{center}

\begin{center}
[Figure 2 about here]
\end{center}

Theorem \ref{thm2} has an immediate application in constructing two combined tests, the Chatterjee-Kendall test and the Chatterjee-quadrant test, defined as
\begin{align*}
\widetilde{I}^{\tau}_{n}(X, Y) & = \max\{|\tau_{n}(X, Y)|, \sqrt{5/2}\xi_{n}(X, Y), \sqrt{5/2}\xi_{n}(Y, X)\}, \\
\widetilde{I}^{Q}_{n}(X, Y) & = \max\{3/2\cdot |Q_{n}(X, Y)|, \sqrt{5/2}\xi_{n}(X, Y), \sqrt{5/2}\xi_{n}(Y, X)\},
\end{align*}
Importantly, both tests admit analytical p-value calculations
\begin{align*}
P(\sqrt{n}\widetilde{I}^{\tau}_{n}(X, Y) > z) & \approx 1 - \Phi^{2}(z) \left[1-2\Phi(-z)\right], \\
P(\sqrt{n}\widetilde{I}^{Q}_{n}(X, Y) > z) & \approx 1 - \Phi^{2}(z) \left[1-2\Phi(-z)\right].
\end{align*}

The performance of the three symmetric tests, $\widetilde{I}^{S}_{n}(X, Y)$, $\widetilde{I}^{\tau}_{n}(X, Y)$ and $\widetilde{I}^{Q}_{n}(X, Y)$, is evaluated in Section 3. Our simulation studies indicate that all tests control the type I error rate effectively. However, the Chatterjee-Kendall test demonstrates the best statistical power across all settings (linear, quadratic, sinusoidal, and stepwise).

\subsection{Extension to the multivariate case}
This section explores the extensions of the proposed tests to the multivariate case. We begin by introducing multivariate versions of $S_{n}(X,~Y)$, $\tau_{n}(X,~Y)$ and $\xi_{n}(X,~Y)$. We omit the multivariate extension of the quadrant correlation due to its low statistical power observed for univariate variables in the simulation studies of Section 3.

Let $\mathbf{X}\in \mathbb{R}^{p}$ and $\mathbf{Y}\in \mathbb{R}^{q}$ be two random vectors, and $(\mathbf{X}_{1},~\mathbf{Y}_{1}),~...,~(\mathbf{X}_{n},~\mathbf{Y}_{n})$ be $n$ $i.i.d$ samples of $(\mathbf{X},~\mathbf{Y})$. The multivariate ranks are defined as
\begin{align*}
R_{i}^{\mathbf{X}} & = \sum_{j=1}^{n} \mathbbm{1}(\mathbf{X}_{j}\leq \mathbf{X}_{i}), \\
R_{i}^{\mathbf{Y}} & = \sum_{j=1}^{n} \mathbbm{1}(\mathbf{Y}_{j}\leq \mathbf{Y}_{i}), \\
R_{i}^{\mathbf{(X,Y)}} & = \sum_{j=1}^{n} \mathbbm{1}(\mathbf{X}_{j}\leq \mathbf{X}_{i},~ \mathbf{Y}_{j}\leq \mathbf{Y}_{i}),
\end{align*}
where $\mathbf{X}_{j}\leq \mathbf{X}_{i}$ denotes element-wise comparison (i.e., all elements in $\mathbf{X}_{j}$ are less than or equal to the corresponding elements in $\mathbf{X}_{i}$. 

Summing the ranks yields
\begin{align*}
R^{\mathbf{X}} & = \sum_{i=1}^{n}R_{i}^{\mathbf{X}},  \\
R^{\mathbf{Y}} & = \sum_{i=1}^{n}R_{i}^{\mathbf{Y}},  \\
R^{\mathbf{(X,Y)}} & = \sum_{i=1}^{n}R_{i}^{\mathbf{(X,Y)}} .
\end{align*}

Grothe et al. (2014) proposed the following multivariate versions of $\tau_{n}$ and $S_{n}$ 
\begin{equation*}
\tau_{n}(\mathbf{X},~\mathbf{Y}) = \frac{n(n-1)R^{\mathbf{(X,Y)}} - R^{\mathbf{X}}R^{\mathbf{Y}}}{\sqrt{ R^{\mathbf{X}}R^{\mathbf{Y}}[n(n-1)-R^{\mathbf{X}}][n(n-1)-R^{\mathbf{Y}}] }},
\end{equation*}
and 
\begin{equation*}
S_{n}(\mathbf{X},~\mathbf{Y}) = \frac{S_{\mathbf{XY}}}{\sqrt{S_{\mathbf{X}}}\sqrt{S_{\mathbf{Y}}}},
\end{equation*}
where
\begin{align*}
S_{\mathbf{XY}} & = \frac{1}{n(n-1)(n-2)}\sum_{i=1}^{n}R_{i}^{\mathbf{X}}R_{i}^{\mathbf{Y}} - \frac{1}{n^{2}(n-1)^2}R^{\mathbf{X}}R^{\mathbf{Y}} - \frac{1}{n(n-1)(n-2)} R^{\mathbf{(X,Y)}}, \\
S_{\mathbf{X}} & = \frac{1}{n(n-1)(n-2)}\sum_{i=1}^{n}R_{i}^{\mathbf{X}}(1-R_{i}^{\mathbf{X}}) -  \frac{1}{n^{2}(n-1)^2}(R^{\mathbf{X}})^2, \\
S_{\mathbf{Y}} & = \frac{1}{n(n-1)(n-2)}\sum_{i=1}^{n}R_{i}^{\mathbf{Y}}(1-R_{i}^{\mathbf{Y}}) -  \frac{1}{n^{2}(n-1)^2}(R^{\mathbf{Y}})^2.
\end{align*}
Grothe et al. (2014) further established the asymptotic normality of $\tau_{n}(\mathbf{X},~\mathbf{Y})$ and $S_{n}(\mathbf{X},~\mathbf{Y})$ by employing the delta method \cite{skvector}.

Azadkia \& Chatterjee (2021) proposed a multivariate extension of Chatterjee's correlation based on nearest neighbors, but this method requires a random scalar for $Y$, thus is not applicable here \cite{azadkia.chatterjee}. Chatterjee (2022) introduced a multivariate extension of $\xi_{n}(\mathbf{X},~\mathbf{Y})$ using Borel isomorphism. This extension transforms a random vector (or the ranks) into a random scalar (preserving its key properties), and works without requiring any distributional assumptions. Let $\eta(\cdot): \mathbb{R}^{p}\rightarrow \mathbb{R}$ be a Borel merging function (Chatterjee gave an example using binary expansion, implemented in R package \textit{XICOR}). As demonstrated by Chatterjee (2022), the binary expansion mapping is a computationally efficient technique that generally produces satisfactory results for Chatterjee's test. A brief description of this method is provided in Appendix A.6. One can define 
\begin{equation*}
\xi_{n}(\mathbf{X},~\mathbf{Y}):=\xi_{n}\left[\eta(\mathbf{X}),~\eta(\mathbf{Y})\right],
\end{equation*}
and show that any result about the univariate case can be transferred to $\xi_{n}(\mathbf{X},~\mathbf{Y})$ \cite{chatterjee.survey}. An immediate result is that for $\mathbf{X}\perp \mathbf{Y}$, $\xi_{n}(\mathbf{X},~\mathbf{Y})\rightarrow N(0,~2/5)$. Therefore we consider the following combinations
\begin{align*}
\widetilde{I}^{S}_{n}(\mathbf{X},~\mathbf{Y}) & = \max(|S_{n}(\mathbf{X},~\mathbf{Y})|/\sigma_{S}, \sqrt{5/2}\xi_{n}(\mathbf{X},~\mathbf{Y}), \sqrt{5/2}\xi_{n}(\mathbf{Y},~\mathbf{X})), \\
\widetilde{I}^{\tau}_{n}(\mathbf{X},~\mathbf{Y}) & = \max(|\tau_{n}(\mathbf{X},~\mathbf{Y})|/\sigma_{\tau}, \sqrt{5/2}\xi_{n}(\mathbf{X},~\mathbf{Y}), \sqrt{5/2}\xi_{n}(\mathbf{Y},~\mathbf{X})),
\end{align*}
where $\sigma_{S}$ and $\sigma_{\tau}$ represent the asymptotic standard deviations of $\sqrt{n}S_{n}(\mathbf{X},~\mathbf{Y})$ and $\sqrt{n}\tau_{n}(\mathbf{X},~\mathbf{Y})$ under independence. It is noteworthy that $\sigma_{S}$ and $\sigma_{\tau}$ depend on the dimensions of $\mathbf{X}$ and $\mathbf{Y}$. Grothe et al. (2014) suggested a gradient-based plug-in method for estimating them. 

Unlike the univariate case, deriving the asymptotic null distribution of the multivariate test statistics is challenging due to different functional forms of $S_{n}(\mathbf{X},~\mathbf{Y})$ (a ratio statistics) and $\xi_{n}(\mathbf{X},~\mathbf{Y})$ (a summation), though simulation study demonstrates that under the null hypothesis, $S_{n}(\mathbf{X},~\mathbf{Y})$ and $\xi_{n}(\mathbf{X},~\mathbf{Y})$ (and $\tau_{n}(\mathbf{X},~\mathbf{Y})$ and $\xi_{n}(\mathbf{X},~\mathbf{Y})$) behave like two independent random variables (see Figures 3 and 4). Here we suggest a permutation test to evaluate the significance. First, we randomly shuffle $(\mathbf{Y}_{1},~...,~\mathbf{Y}_{n})$ for $B$ times. For each permutation, we calculate $\widetilde{I}^{S}_{n}(\mathbf{X},~\mathbf{Y})$ and $\widetilde{I}^{\tau}_{n}(\mathbf{X},~\mathbf{Y})$ based on the shuffled data. The resulting $\widetilde{I}^{S}_{n}(\mathbf{X},~\mathbf{Y})$ and $\widetilde{I}^{\tau}_{n}(\mathbf{X},~\mathbf{Y})$ are used to approximate the null distributions and estimate the p-values. An additional benefit of permutation tests is that they can directly estimate the unknown standard deviations $\sigma_{S}$ and $\sigma_{\tau}$, from the shuffled data.

\begin{center}
[Figure 3 about here]
\end{center}

\begin{center}
[Figure 4 about here]
\end{center}

Alternatively, one may consider a simpler form of the multivariate Kendall and Spearman correlations, instead of those introduced by \cite{skvector}. Recall that under Borel isomorphism and $\mathbf{X}\perp \mathbf{Y}$, $[\eta(\mathbf{X}),~\eta(\mathbf{Y})]$ become independent random scalars, thus all the previously established asymptotic results for the univariate case directly apply to the multivariate case. For instance, we have

$$\sqrt{n}\begin{bmatrix}
\tau_{n}\left[\eta(\mathbf{X}),~\eta(\mathbf{Y})\right]\\
\xi_{n}\left[\eta(\mathbf{X}),~\eta(\mathbf{Y})\right]\\
\xi_{n}\left[\eta(\mathbf{Y}),~\eta(\mathbf{X})\right]
\end{bmatrix}
\xrightarrow{d} N\left [
\begin{pmatrix}
0\\
0\\
0
\end{pmatrix}, 
\begin{pmatrix}
4/9 & 0 & 0 \\
0 & 2/5 & 0 \\
0 & 0 & 2/5
\end{pmatrix} \right ],$$ and 
$$\sqrt{n}\begin{bmatrix}
S_{n}\left[\eta(\mathbf{X}),~\eta(\mathbf{Y})\right]\\
\xi_{n}\left[\eta(\mathbf{X}),~\eta(\mathbf{Y})\right]\\
\xi_{n}\left[\eta(\mathbf{Y}),~\eta(\mathbf{X})\right]
\end{bmatrix}
\xrightarrow{d} N\left [
\begin{pmatrix}
0\\
0\\
0
\end{pmatrix}, 
\begin{pmatrix}
1 & 0 & 0 \\
0 & 2/5 & 0 \\
0 & 0 & 2/5
\end{pmatrix} \right ],$$
as $n\rightarrow\infty$. 

While the max-type test statistic can be constructed using the same formula as in the univariate case, our simulations in Section 3 suggest that the Borel isomorphism approach might not be ideal for Spearman's and Kendall's correlations. In fact, these tests may have extremely low statistical power, particularly for detecting monotonic relationships. Therefore we recommend the first approach employing permutation tests based on Grothe et al.'s formulas.

\section{Simulation studies}
To evaluate the performance of the proposed tests, we conducted two simulation studies. The first focused on the univariate case, comparing the empirical power of seven independence tests: symmetrized Chatterjee ($\xi^{sym}_{n}$), Chatterjee-Spearman ($\widetilde{I}^{S}_{n}$), Chatterjee-Kendall ($\widetilde{I}^{\tau}_{n}$), Chatterjee-quadrant ($\widetilde{I}^{Q}_{n}$), Hoeffding's $D$, Blum-Kiefer-Rosenblatt's $R$, and Bergsma-Dassios' $\tau^*$, under sample sizes $\{20, 40, 60, 80, 100\}$. The symmetrized Chatterjee correlation is defined as $\xi^{sym}_{n} = \max\left\{ \sqrt{5/2}\xi_{n}(X, Y),  \sqrt{5/2}\xi_{n}(Y, X)\right\}$, and its asymptotic p-value can be calculated as $P(\sqrt{n}\xi^{sym}_{n}(X, Y) > z) \approx 1 - \Phi^{2}(z)$, where $\Phi(\cdot)$ represents the $c.d.f.$ of the standard normal distribution. The calculations of $D_{n}$, $R_{n}$ and $\tau^{*}_{n}$ were performed using R package \textit{independence}. The following four alternatives were considered, where $X\sim \mbox{Uniform}[-1, 1]$, $Z\sim N(0,1)$ and $Z\perp X$ (to investigate whether our proposed method is applicable to discontinuous distributions, we specifically added a stepwise setting)

\begin{itemize}
\item[1.] Linear: $Y = X+Z$
\item[2.] Quadratic:  $Y = X^2+0.3Z$.
\item[3.] Stepwise (monotonic):  $Y = \mathbbm{1}_{\{-1\leq X\leq -0.5\}}+2*\mathbbm{1}_{\{-0.5<X\leq0\}}+3*\mathbbm{1}_{\{0<X\leq 0.5\}}+4*\mathbbm{1}_{\{0.5<X\leq 1\}}+2Z$.
\item[4.] Sinusoid:  $Y = \cos(2\pi X)+0.75Z$.
\end{itemize} 

Table 1 summarizes the empirical power of the tests based on $5,000$ simulation runs at the significance level of 0.05. First, we observe that Hoeffding's $D$, Blum-Kiefer-Rosenblatt's $R$, and Bergsma-Dassios' $\tau^*$ show similar power across all settings. In particular, these tests are powerful for the linear, quadratic and stepwise settings, but not for the sinusoidal setting. The symmetrized Chatterjee test has high power for the two non-monotonic settings (quadratic and sinusoidal), but performs poorly in the linear and stepwise settings. For the three combined tests that we proposed, Chatterjee-Spearman and Chatterjee-Kendall outperform Chatterjee-quadrant especially for the two monotonic settings (linear and stepwise). For instance, in the stepwise setting with $n=40$, Chatterjee-Kendall and Chatterjee-Spearman achieve a power of 0.827 and 0.809, respectively, compared to 0.527 for Chatterjee-quadrant. In most settings, Chatterjee-Kendall and Chatterjee-Spearman perform comparably. However, for relatively small sample sizes, Chatterjee-Kendall is more powerful than Chatterjee-Spearman. For instance, in the linear and stepwise settings with $n=20$, the power of Chatterjee-Kendall is about 5-6\% higher than Chatterjee-Spearman. A plausible interpretation for this is that both $\widetilde{I}^{S}_{n}$ and $\widetilde{I}^{\tau}_{n}$ incorporate Chatterjee's correlation, which is consistent against all alternatives. Chatterjee's correlation generally offers satisfactory power under low noise conditions and sufficient sample sizes, regardless of the specific correlation pattern (linear or nonlinear). However, its power can diminish in the presence of significant noise or relatively small sample sizes. When Chatterjee's correlation fails to capture the dependence structure in both tests, it typically indicates noisy data and/or limited sample size. In such scenarios, Kendall's correlation has been demonstrated to be more robust and powerful compared to Spearman's correlation (see \cite{crouxdehon, KS1}). Consequently, the Chatterjee-Kendall test demonstrates better overall power compared to the Chatterjee-Spearman test.

Table 2 summarizes the empirical size of the tests over $50,000$ simulation runs, where $X\perp Y$, $X\sim\mbox{Uniform}[-1, 1]$ and $Y\sim N(0,1)$. It can be seen that all seven tests control the Type I error rate at the nominal level of $0.05$. Some of tests are, however, slightly conservative for small sample sizes (0.037 for $\xi_{n}$ and 0.038 for $(\xi_{n}, S_{n})$ under $n=20$). 

Our second simulation study investigates the performance of the multivariate tests (Section 2.3), including symmetrized Chatterjee ($\xi^{sym}_{n}$), Spearman ($S_{n}$), Kendall ($\tau_{n}$), Chatterjee-Spearman ($\widetilde{I}^{S}_{n}$), and Chatterjee-Kendall ($\widetilde{I}^{\tau}_{n}$). We evaluate two approaches for multivariate versions of $S_{n}$ and $\tau_{n}$: one based on Grothe et al.'s formulas and the other based on Borel isomorphism. For Grothe et al.'s method, the p-values are calculated using $5,000$ random permutations. For Borel isomorphism method, we calculated both analytical and permutation p-values. The following six alternatives were considered, where $\mathbf{X} = (X_{[1]},~X_{[2]},~X_{[3]}),~\mathbf{Y} = (Y_{[1]},~Y_{[2]},~Y_{[3]})$, $Z_{[i]}\sim N(0,1)$, $i=1,~2,~3.$
\begin{itemize}
\item[1.] $X_{[i]} \sim \mbox{Uniform}[0, 1]$, $i = 1,~2,~3,$ and 
$$
\begin{cases}
Y_{[1]} & =  X_{[1]} - X_{[2]} + 2X_{[3]} + 2Z_{[1]}, \\
Y_{[2]} & =  -X_{[1]} + 3X_{[2]} -\frac{1}{2}X_{[3]} + 2Z_{[2]}, \\
Y_{[3]} & =  \frac{3}{2}X_{[1]} + X_{[2]} + 2X_{[3]} + 2Z_{[3]}.
\end{cases}
$$
\item[2.] $X_{[i]} \sim \mbox{Uniform}[-1, 1]$, $i = 1,~2,~3,$ and 
$$
\begin{cases}
Y_{[1]} & =  3X_{[1]} + 4Z_{[1]}, \\
Y_{[2]} & =  X_{[2]} + 2Z_{[2]}, \\
Y_{[3]} & =  2X_{[3]} + 3Z_{[3]}.
\end{cases}
$$
\item[3.] $X_{[i]} \sim \mbox{Uniform}[0, 2]$, $i = 1,~2,~3,$ and 
$$
\begin{cases}
Y_{[1]} & =  X_{[2]} + 2X_{[3]} + 4Z_{[1]}, \\
Y_{[2]} & =  2X_{[1]} + \frac{1}{2}X_{[3]} + 4Z_{[2]}, \\
Y_{[3]} & =  X_{[1]} + 3X_{[2]} + 4Z_{[3]}.
\end{cases}
$$
\item[4.] $X_{[i]} \sim \mbox{Uniform}[-2, 2]$, $i = 1,~2,~3,$ and 
$$
\begin{cases}
Y_{[1]} & =  2X^{2}_{[1]} + 4|X_{[2]}| + \cos(2\pi X_{[3]}) + 2Z_{[1]}, \\
Y_{[2]} & =  2\cos(2\pi X_{[1]}) + 3X^{2}_{[2]} + |X_{[3]}| + 2Z_{[2]}, \\
Y_{[3]} & =  3|X_{[1]}| + 2 \cos(2\pi X_{[2]}) + 2X^{2}_{[3]} + 2Z_{[3]}.
\end{cases}
$$
\item[5.] $X_{[i]} \sim \mbox{Uniform}[-2, 2]$, $i = 1,~2,~3,$ and  
$$
\begin{cases}
Y_{[1]} & =  2X^{2}_{[1]} + 5Z_{[1]}, \\
Y_{[2]} & =  4X^{2}_{[2]} + 5Z_{[2]}, \\
Y_{[3]} & =  6X^{2}_{[3]} + 5Z_{[3]}.
\end{cases}
$$
\item[6.] $X_{[i]} \sim \mbox{Uniform}[-1, 1]$, $i = 1,~2,~3,$ and 
$$
\begin{cases}
Y_{[1]} & =  \cos(2\pi X_{[1]})  + Z_{[1]}/2, \\
Y_{[2]} & =  \cos(4\pi X_{[2]}) + Z_{[2]}/2, \\
Y_{[3]} & =  \cos(6\pi X_{[3]}) + Z_{[3]}/2.
\end{cases}
$$
\end{itemize} 

Table 3 summarizes the empirical power of the five tests, where Grothe et al.'s formula is used for the multivariate versions of $S_{n}$ and $\tau_{n}$. The results are consistent with the univariate case: the symmetrized Chatterjee correlation remains the most powerful test in the nonlinear settings (4, 5 and 6) but the least powerful in the linear settings (1, 2 and 3). In contrast, Spearman's and Kendall's tests are powerful in the linear settings, but suffering from extremely low power in the nonlinear settings. Notably, in the linear settings, Kendall's test is substantially more powerful than Spearman's test (10-23\% difference). The Chatterjee-Kendall test has the overall best power for all settings. All the five tests control the type I error rate as they are permutation-based. 

Table 4 summarizes the empirical power of the five tests, where the multivariate $S_{n}$ and $\tau_{n}$ are computed using Borel isomorphism. Interestingly, Kendall's correlation outperforms the others in several linear and nonlinear settings (for both analytical and permutation p-values). This might seem counterintuitive as Kendall's test typically measures monotonic associations. However, in the linear settings, all the tests exhibit low statistical power. These findings suggest that under Borel isomorphism, combining $\xi_{n}$ with $S_{n}$ or $\tau_{n}$ might not be ideal for multivariate data, unlike the complementary behavior observed in the univariate case. For Kendall's and Spearman's tests based on Borel isomorphism, the use of analytical p-values tend to over-estimate the power in certain settings. In contrast, the symmetrized Chatterjee test with analytical p-values is consistently more conservative than the permutation test in all settings. Similar findings are also reported in \cite{Zhang23} (Figures 3 and 4). Table 5 presents the empirical size of the tests based on Borel isomorphism, where $X_{[i]} \sim \mbox{Uniform}[-1, 1]$, and $Y_{[i]}\sim N(0,1)$, $i = 1,~2,~3.$ It can be seen that for several tests, including multivariate Kendall, Spearman and Chatterjee-Kendall, the use of analytical p-values often fail to control the type I error rate, especially for smaller sample sizes. For instance, for $n=20$, Kendall's test shows an empirical type I error rate of 0.078 (nominal 0.05). Given these limitations, we recommend the permutation tests based on Grothe et al.'s formulas for practical applications, which offer better overall power.

\section{Real data applications}
\subsection{Yeast transcriptomics data}
To assess our proposed tests, we utilized a transcriptomics dataset from \cite{spellman}. This dataset originally contained expression levels of 6,223 yeast genes measured at 23 time points during the cell cycle. After the preprocessing by \cite{mic} to remove genes with missing data, the final dataset comprised 4,381 genes, which is readily available through the R package \textit{minerva}). Given its widespread use in the literature, including in Chatterjee (2021), this dataset provides a robust benchmark for evaluating correlation measures.

We analyzed this data using five tests, including the symmetrized Chatterjee ($\xi^{sym}_{n}$), Spearman ($S_{n}$), Kendall ($\tau_{n}$), Chatterjee-Spearman ($\widetilde{I}^{S}_{n}$), and Chatterjee-Kendall ($\widetilde{I}^{\tau}_{n}$). For all five methods, p-values were calculated asymptotically and adjusted using the Benjamini-Hochberg procedure to control the false discovery rate (FDR) at 0.05. Figure 5 summarizes the number of significant genes identified by each test. The Chatterjee-Kendall test detected the most genes (897) that exhibit cell cycle-related expression changes, outperforming individual Kendall (614) and Chatterjee (624) tests. This is attributed to its ability to capture diverse expression patterns, including both monotonic and non-monotonic (e.g., oscillatory) changes. Figures 6 and 7 illustrate some random examples of genes identified by the Chatterjee-Kendall test but missed by Kendall's test and Chatterjee's test, respectively. The genes in Figure 6 show oscillatory patterns, while those in Figure 7 show smoother trends. Figure 8 presents some random examples of genes identified by Chatterjee-Kendall but missed by Chatterjee-Spearman. These genes exhibit noisy expression profiles, highlighting the better robustness of Kendall's test to outliers. This analysis demonstrates the enhanced sensitivity of the combined tests in detecting various correlation patterns. For noisy data, the Chatterjee-Kendall test is particularly advantageous due to its robustness.

\begin{center}
[Figure 5 about here]
\end{center}
\begin{center}
[Figure 6 about here]
\end{center}
\begin{center}
[Figure 7 about here]
\end{center}
\begin{center}
[Figure 8 about here]
\end{center}

\subsection{Salespeople data}
We applied our multivariate tests to a real-world dataset from a company that surveyed 50 employees to understand the factors influencing sales performance (Table 9.12 in \cite{salespeople}). We focused on the multivariate association between two sets of variables: (1) sales performance metrics (growth, profitability, and new accounts) and (2) reasoning test scores (mechanical and abstract). Given the potential issues with the Borel isomorphism for Spearman's and Kendall's tests, as discussed in our simulation studies, we used Grothe et al.'s formula and calculated p-values using 10,000 random permutations. All five multivariate tests indicated a highly significant association between sales performance and reasoning test scores (all p-values $<2\times 10^{-4}$).

To assess the performance of these tests under smaller sample sizes, we conducted a simulation study. We randomly selected 15 samples from the original dataset, repeated this process 5,000 times, and compared the proportion of times each test rejected the null hypothesis. The symmetrized Chatterjee's test exhibited the lowest sensitivity (0.653) to smaller sample sizes. In contrast, the Kendall and Chatterjee-Kendall tests demonstrated the highest sensitivity (0.958 and 0.951, respectively). The Spearman and Chatterjee-Spearman tests showed intermediate sensitivity (0.863 and 0.867, respectively). This real-world example further underscores the good performance of our Chatterjee-Kendall test (based on Grothe et al.'s formula) in practical settings.

\section{Discussion and conclusions}
Chatterjee's correlation has gained significant interest due to its simple form and appealing statistical properties. The only one disadvantage is its inferior performance in testing monotonic associations. Our prior work addressed this by combining it with Spearman's correlation, which is powerful in detecting monotonic associations. This work delves into three key extensions of the Chatterjee-Spearman correlation. Firstly, we examine the symmetrized version of the statistic, along with its asymptotic distribution. This eliminates the need to designate response and independent variables in practical applications, allowing for direct application to diverse datasets, such as gene co-expression data. Secondly, we demonstrate the adaptability of this framework by showing that Spearman's correlation can be substituted with other rank-based measures like Kendall's $\tau$ or quadrant correlation. This is because Chatterjee's correlation and these alternatives are also asymptotically normal and independent under the null hypothesis. Simulation studies demonstrate the competitiveness of the symmetrized Chatterjee-Spearman and Chatterjee-Kendall correlations with existing tests. Notably, the Chatterjee-Kendall correlation exhibits slightly better power, particularly for smaller samples. Thirdly, we explore how these new tests can be adapted for analyzing multivariate data. These extensions significantly broaden the applicability of this method, making it suitable for a wider range of scenarios.

The methods introduced in this paper have some limitations. The multivariate extensions based on Grothe et al.'s formulas are currently restricted to low dimensions because they rely on multivariate ranks $\{R_{i}^{\mathbf{X}},~R_{i}^{\mathbf{Y}},~R_{i}^{\mathbf{(X,Y)}} \}$, which can be zeros or very sparse under high dimensions ($\min(p,~q) \gg n$), leading to extremely low testing power. For high-dimensional independence testing, established methods such as the distance correlation t-test by Sz\'{e}kely \& Rizzo (2013) are preferable \cite{dc, dc1, dc2, dc3}. 

Our combined tests, including the Chatterjee-Spearman and Chatterjee-Kendall tests, utilize the maximum of individual statistics. Enlightened by a reviewer's comments, it would be interesting to investigate other potential combinations, such as weighted averages or the $L_p$ norms of individual statistics. However, these methods involve additional parameters, such as weights or the order $p$, which would require substantial theoretical and simulation work to optimize. We therefore leave this for future research.

While our simulation studies suggest superior power for the Chatterjee-Kendall test compared to existing rank correlations, there is a lack of theoretical justification. By investigating their power against local rotation and mixture alternatives, Shi et al. (2021) showed that Chatterjee's correlation is unfortunately sub-optimal compared to Hoeffding's $D$, Blum-Kiefer-Rosenblatt's $R$, and Bergsma-Dassios' $\tau^*$ \cite{ShiHan21}. An important question is that if the Chatterjee-Kendall and Chatterjee-Spearman tests are also sub-optimal compared to $D$, $R$ and $\tau^*$ under certain local alternatives. This is a challenging question that deserves further investigation, and we leave it for future work.

\section*{Data availability}
\noindent
The yeast transcriptomics data is available in the R package \textit{minerva} at \url{cran.r-project.org/web/packages/minerva}. The salespeople data can be found in Tables 9.12 of \textit{Applied Multivariate Statistical Analysis} (Johnson \& Wichern, 2002, 6th edition).

\section*{Acknowledgement}
\noindent
The work was supported by an NSF DBI Biology Integration Institute (BII) Grant (Award No. 2119968; PI-Ceballos).

 \section*{Competing Interests}
\noindent
The author has declared that no competing interests exist.

\section*{Appendix}
\subsection*{A.1. Proof of Theorem 1}
We begin with some necessary notations. For data rearranged with respect to $X$, i.e., $\{(X_{(1)},~Y_{X(1)}),~...,~(X_{(n)},~Y_{X(n)})\}$, where $X_{(1)}< ... < X_{(n)}$ and $\{Y_{X(1)},~...,~Y_{X(n)}\}$ denote the concomitants, let $R(Y_{X(i)})$ be the rank of $Y_{X(i)}$, i.e., $R(Y_{X(i)}) = \sum_{j=1}^{n}\mathbbm{1}\{Y_{(j)}\leq Y_{X(i)}\}$. On the other hand, $\{X_{Y(1)},~...,~X_{Y(n)}\}$ denote the concomitants, and $R(X_{Y(i)})$ is the rank of $X_{Y(i)}$. Then the two Chatterjee correlations and Spearman correlation can be expressed as
\begin{align*}
\xi_{n}(X, Y) & = 1-\frac{3}{n^2-1}\sum_{i=1}^{n-1}|R(Y_{X(i+1)})-R(Y_{X(i)})|, \\
\xi_{n}(Y, X) & = 1-\frac{3}{n^2-1}\sum_{i=1}^{n-1}|R(X_{Y(i+1)})-R(X_{Y(i)})|, \\
S_{n}(X, Y) & = 1 - \frac{6\sum_{i=1}^{n}[i-R(Y_{X(i)})]^2}{n(n^2-1)} = - \frac{3(n+1)}{n-1} + \frac{12\sum_{i=1}^{n}iR(Y_{X(i)})}{n(n^2-1)}.
\end{align*}
Under independence, we have $\lim_{n\rightarrow\infty}V[\sqrt{n}S_{n}(X, Y)] = 1$ and $\lim_{n\rightarrow\infty}V[\sqrt{n}\xi_{n}(X, Y)] = \lim_{n\rightarrow\infty}V[\sqrt{n}\xi_{n}(Y, X)]  = 2/5$. Also, by Lemmas \ref{lem1} and \ref{lem2}, we have 
\begin{align*}
\lim_{n\rightarrow\infty} Cov\left[ \sqrt{n}S_{n}(X, Y),~ \sqrt{n}\xi_{n}(X, Y) \right] & = 0, \\
\lim_{n\rightarrow\infty} Cov\left[ \sqrt{n}S_{n}(X, Y),~ \sqrt{n}\xi_{n}(Y, X) \right] & = 0, \\
\lim_{n\rightarrow\infty} Cov\left[ \sqrt{n}\xi_{n}(X, Y),~ \sqrt{n}\xi_{n}(Y, X) \right] & = 0.
\end{align*}
Thus we only need to show the joint normality of $[\sqrt{n}S_{n}(X, Y),~\sqrt{n}\xi_{n}(X, Y),~\sqrt{n}\xi_{n}(Y, X)]$. Our proof is a generalization of the proofs for Lemma 3 in \cite{Zhang23} and Lemma 2 in \cite{Zhang24}. 

Let $F_{X}(x)$ and $F_{Y}(y)$ be the $c.d.f.$'s of $X$ and $Y$, $\widehat{F}_{X}(x)$ and $\widehat{F}_{Y}(y)$ be the empirical $c.d.f.$'s, i.e.,
\begin{align*}
\widehat{F}_{X}(x) & = \frac{1}{n}\sum_{i=1}^{n}\mathbbm{1}\{X_{i}\leq x\}, \\
\widehat{F}_{Y}(y) & = \frac{1}{n}\sum_{i=1}^{n}\mathbbm{1}\{Y_{i}\leq y\},
\end{align*}
and $\widehat{F}_{X}(X_{Y(i)}) = R(X_{Y(i)})/n$. In addition, we define $U_{i}:= F_{X}(X_{Y(i)})$ and $V_{i} := F_{Y}(Y_{X(i)})$. It is straightforward that under independence of $X$ and $Y$, $(U_{1},~...,~U_{n})$ are $i.i.d.$ uniform on $[0,~1]$. However, it should be noted that $(U_{1},~...,~U_{n})$ and $(V_{1},~...,~V_{n})$ are generally dependent. For $\xi_{n}$, using Equations 5-8 in \cite{angus}, we have 
\begin{align*}
\frac{\sum_{i=1}^{n-1}|R(Y_{X(i+1)})-R(Y_{X(i)})|-n^2/3}{n\sqrt{n}} & \stackrel{d}{\approx} \frac{1}{\sqrt{n}}\sum_{i=1}^{n-1} \left[ |U_{i+1}-U_{i}| + 2U_{i}(1-U_{i}) - \frac{2}{3} \right], \\
\frac{\sum_{i=1}^{n-1}|R(X_{Y(i+1)})-R(X_{Y(i)})|-n^2/3}{n\sqrt{n}} & \stackrel{d}{\approx} \frac{1}{\sqrt{n}}\sum_{i=1}^{n-1} \left[ |V_{i+1}-V_{i}| + 2V_{i}(1-V_{i}) - \frac{2}{3} \right], 
\end{align*}
where $\stackrel{d}{\approx}$ represents asymptotic equivalence in distribution. For $S_{n}$, from the proof of Lemma 2 in \cite{Zhang24}, we have
\begin{equation*}
\frac{\sum_{i=1}^{n}iR(Y_{X(i)})-n^3/4}{n^3} \stackrel{d}{\approx} \frac{1}{\sqrt{n}}\sum_{i=1}^{n}\left[ \frac{i}{n+1} - \frac{1}{2} \right]U_{i}.
\end{equation*}
Recall that $U_{i}:= F_{X}(X_{Y(i)})$ and $V_{i} := F_{Y}(Y_{X(i)})$, it suffices to show that for any constants $a,~b,~c\in \mathbb{R}$, the following quantity converges to a normal distribution
\begin{align*}
J_{n} = \frac{1}{\sqrt{n}} \sum_{i=1}^{n} & \bigg[ a|F_{X}(X_{i})-F_{X}(X_{N_{Y}(i)})| + 2aF_{X}(X_{i})(1-F_{X}(X_{i}))  \\
& + b|F_{Y}(Y_{i})-F_{Y}(Y_{N_{X}(i)})| + 2bF_{Y}(Y_{i})(1-F_{Y}(Y_{i})) \\
& + c\left( \frac{i}{n+1} - \frac{1}{2} \right)F_{X}(X_{i}) + \frac{2(a+b)}{3} \bigg],
\end{align*}
where $(X_{N_{X}(i)},~Y_{N_{X}(i)})$ represents the right nearest neighbor of $(X_{i},~Y_{i})$ in terms of $X$ (define $N_{X}(i) = 1$ if $X_{i} = X_{(n)}$). Here, $J_{n}$ is the sum of dependent variables, and we leverage Chatterjee's central limit theorem based on interaction graphs \cite{chatterjee.clt, auddy}. Motivated by \cite{auddy}, we define a graphical rule $\mathcal{G}$ using $J_{n}$, which we will show to be an interaction rule (Section 2.3 of \cite{chatterjee.clt}, page 5). Let $\mathcal{M} :=\{(X_{1}, Y_{1}), ..., (X_{n}, Y_{n})\}$ and $\mathcal{M}' := \{(X'_{1}, Y'_{1}), ..., (X'_{n}, Y'_{n})\}$ be two $i.i.d.$ samples and 
\begin{align*}
\mathcal{M}^{i} & :=\{(X_{1}, Y_{1}), ..., (X'_{i}, Y'_{i}), ..., (X_{n}, Y_{n})\},  \\
\mathcal{M}^{j} & :=\{(X_{1}, Y_{1}), ..., (X'_{j}, Y'_{j}), ..., (X_{n}, Y_{n})\},  \\
\mathcal{M}^{ij} & :=\{(X_{1}, Y_{1}), ..., (X'_{i}, Y'_{i}), ..., (X'_{j}, Y'_{j}), ..., (X_{n}, Y_{n})\}. 
\end{align*}
Following Equation 4.17 of \cite{auddy}, we define $\mathcal{G}(\mathcal{M})$ on $\{1, 2, ..., n\}$. Let
$$
D^{X}_{\mathcal{M}}(i, j)=
\begin{cases}
\infty, ~~\mbox{if} ~X_{i}>X_{j}\\
\#\{l: X_{i}<X_{l}<X_{j}\}, ~~\mbox{if}~X_{i}<X_{j},
\end{cases}
$$
and
$$
D^{Y}_{\mathcal{M}}(i, j)=
\begin{cases}
\infty, ~~\mbox{if} ~Y_{i}>Y_{j}\\
\#\{l: Y_{i}<Y_{l}<Y_{j}\}, ~~\mbox{if}~Y_{i}<Y_{j}.
\end{cases}
$$
For a pair of indices $\{i, j\}$, there is an edge between them if there exists an $l \in \{1, 2, ..., n\}$, such that
\begin{align*}
D^{X}_{\mathcal{M}}(l, i)\leq 2~&\mbox{and}~D^{X}_{\mathcal{M}}(l, j)\leq 2 \\
& \mbox{or}\\
D^{Y}_{\mathcal{M}}(l, i)\leq 2~&\mbox{and}~D^{Y}_{\mathcal{M}}(l, j)\leq 2.
\end{align*}
It is obvious that this rule is invariant under relabeling of indices, therefore it is symmetric (see the definition of symmetric rules in \cite{chatterjee.clt}, page 5). The statistic $J_{n}(\mathcal{M})$ can be decomposed into three parts
\begin{align*}
J_{n, 1}(\mathcal{M}) & = \frac{1}{\sqrt{n}}\sum_{i=1}^{n}\left[ a|F_{X}(X_{i})-F_{X}(X_{N_{Y}(i)})| \right], \\
J_{n, 2}(\mathcal{M}) & = \frac{1}{\sqrt{n}}\sum_{i=1}^{n}\left[ b|F_{Y}(Y_{i})-F_{Y}(Y_{N_{X}(i)})| \right], \\
J_{n, 3}(\mathcal{M}) & = \frac{1}{\sqrt{n}}\sum_{i=1}^{n}\left[ 2aF_{X}(X_{i})(1-F_{X}(X_{i})) + 2bF_{Y}(Y_{i})(1-F_{Y}(Y_{i})) + c\left( \frac{i}{n+1} - \frac{1}{2} \right)F_{X}(X_{i}) + \frac{2(a+b)}{3} \right ].
\end{align*}

We also need to show that $\mathcal{G}(\mathcal{M})$ is an interaction rule. For any $\{i, j\}$, if there is no edge between them, there does not exist an $l\in\{1, 2, ..., n\}$, such that $D^{X}_{\mathcal{M}}(l, i)\leq 2$ and $D^{X}_{\mathcal{M}}(l, j)\leq 2$. Following Lemma 4 of \cite{auddy}, and Lemma 3 of \cite{Zhang23}
\begin{align*}
J_{n, 1}(\mathcal{M})-J_{n, 1}(\mathcal{M}^{i})-J_{n, 1}(\mathcal{M}^{j})+J_{n, 1}(\mathcal{M}^{ij}) & = 0, \\
J_{n, 2}(\mathcal{M})-J_{n, 2}(\mathcal{M}^{i})-J_{n, 2}(\mathcal{M}^{j})+J_{n, 2}(\mathcal{M}^{ij}) & = 0.
\end{align*}
The third part, $J_{n, 3}$, does not involve nearest neighbor, and it can be easily verified that $J_{n, 3}(\mathcal{M})-J_{n, 3}(\mathcal{M}^{i})-J_{n, 3}(\mathcal{M}^{j})+J_{n, 3}(\mathcal{M}^{ij}) = 0$. Therefore $J_{n}(\mathcal{M})-J_{n}(\mathcal{M}^{i})-J_{n}(\mathcal{M}^{j})+J_{n}(\mathcal{M}^{ij}) = 0$, i.e., $\mathcal{G}$ is a symmetric interaction rule. The extended graph $\mathcal{G}'$ on $\{1, 2, ..., n+4\}$ can be constructed same as in Lemma 3 of \cite{Zhang23}. Using Theorem 2.5 of \cite{chatterjee.clt}, there exists a constant $C>0$, such that
$$\mathcal{D}(J_{n}) \leq \frac{C}{\sqrt{n}\sigma^2} + \frac{C}{2\sqrt{n}\sigma^3},$$
where $\sigma^{2} = \mbox{V}(J_{n})$, and $\mathcal{D}(J_{n})$ is the Wasserstein distance between $(J_{n}-E(J_{n}))/\sigma$ and $N(0 ,1)$.

We now derive the variance term $\sigma^{2}$. Let $d_{i} = i/(n+1) - 1/2$, we have 
\begin{align}
\sigma^{2}  = & V\left\{\frac{1}{\sqrt{n}}\sum_{i=1}^{n}\left[ a|U_{i+1}-U_{i}| + b|V_{i+1}-V_{i}| + 2aU_{i}(1-U_{i}) + cd_{i}U_{i} + 2bV_{i}(1-V_{i})\right]\right\} \nonumber \\
 = & V\left\{\frac{1}{\sqrt{n}}\sum_{i=1}^{n}[ a|U_{i+1}-U_{i}| + 2aU_{i}(1-U_{i})]\right\} + V\left\{\frac{1}{\sqrt{n}}\sum_{i=1}^{n}[ b|V_{i+1}-V_{i}| + 2bV_{i}(1-V_{i})]\right\} \label{pa1} \\
& + Cov\left\{\frac{1}{\sqrt{n}}\sum_{i=1}^{n}[ a|U_{i+1}-U_{i}| + 2aU_{i}(1-U_{i})],~ \frac{1}{\sqrt{n}}\sum_{i=1}^{n}[ b|V_{i+1}-V_{i}| + 2bV_{i}(1-V_{i})]\right\} \label{pa2}\\
& + Cov\left\{\frac{1}{\sqrt{n}}\sum_{i=1}^{n}[ a|U_{i+1}-U_{i}| + 2aU_{i}(1-U_{i})],~ \frac{1}{\sqrt{n}}\sum_{i=1}^{n}cd_{i}U_{i}\right\},\label{pa3} \\
& + Cov\left\{\frac{1}{\sqrt{n}}\sum_{i=1}^{n}[ b|V_{i+1}-V_{i}| + 2bV_{i}(1-V_{i})],~ \frac{1}{\sqrt{n}}\sum_{i=1}^{n}cd_{i}U_{i}\right\},\label{pa4} \\
& + V\left(\frac{1}{\sqrt{n}}\sum_{i=1}^{n}cd_{i}U_{i}\right). \label{pa5}
\end{align}
From Equation (14) in \cite{angus}, term (\ref{pa1}) is $2(a^2+b^2)/45 + O(1/n)$. From Lemma 3 of \cite{Zhang23}, term (\ref{pa2}) is $O(1/n)$. Using the facts that for two independent variables $(U_{1},~U_{2})\sim \mbox{Uniform}(0,~ 1)$, $V(U_{1}) = 1/12$, $Cov(U_{1},~U_{1}^2) = 1/12$, and $Cov(|U_{1}-U_{2}|,~U_{1}) = 0$, it can be shown that  
$$Cov\left\{\frac{1}{\sqrt{n}}\sum_{i=1}^{n}[ a|U_{i+1}-U_{i}| + 2aU_{i}(1-U_{i})],~ \frac{1}{\sqrt{n}}\sum_{i=1}^{n}cd_{i}U_{i}\right\}  = 0,$$
and 
$$V\left(\frac{1}{\sqrt{n}}\sum_{i=1}^{n}cd_{i}U_{i}\right) = c^2\left(\frac{2n+1}{72n+1}-\frac{1}{48} \right)= \frac{c^2}{144} +O\left( \frac{1}{n} \right).$$
Term (\ref{pa4}) is a covariance, therefore it is bounded by 
$$ \frac{1}{2}V\left(\frac{1}{\sqrt{n}}\sum_{i=1}^{n}cd_{i}U_{i}\right) + \frac{1}{2} V\left\{\frac{1}{\sqrt{n}}\sum_{i=1}^{n}[ b|V_{i+1}-V_{i}| + 2bV_{i}(1-V_{i})]\right\} = \frac{b^2}{45}+ \frac{c^2}{288} + O\left( \frac{1}{n} \right).$$
Therefore we have 
$$\frac{2a^2}{45}+ \frac{b^2}{45}+ \frac{c^2}{288} + O\left( \frac{1}{n} \right) \leq \sigma^{2} \leq \frac{2a^2}{45} + \frac{b^2}{15} + \frac{c^2}{96} +O\left( \frac{1}{n} \right), $$
and the Wasserstein distance between $(J_{n}-E(J_{n}))/\sigma$ and $N(0 ,1)$ converges to 0, which completes the proof.

\subsection*{A.2. Proof of Lemma \ref{lem2}}
First, we rewrite the quadrant correlation as
\begin{equation*}
Q_{n}(X, Y) = \frac{1}{n}\sum_{i=1}^{n}Z_{i},
\end{equation*}
where
\begin{equation*}
Z_{i}=\mbox{sgn}[(X_{i} - \mbox{med}_{n}(X))(Y_{i} - \mbox{med}_{n}(Y))].
\end{equation*}

We now derive $V(Z_{i})$ and $Cov(Z_{i},~Z_{j})$. Due to the difference in how the sample median is defined for odd and even sample sizes, the expressions of variance and covariance will also differ. Therefore, we discuss the two cases separately. When $n$ is even, we have $V(Z_{i}) = E(Z_{i}^2) = 1$ and $Cov(Z_{i},~Z_{j}) = E(Z_{i}Z_{j})$. The expectation $E(Z_{i}Z_{j})$ can be decomposed as follows
\begin{align*}
E(Z_{i}Z_{j}) = &~ E[Z_{i}Z_{j}|X_{i}<\mbox{med}_{n}(X),~X_{j}<\mbox{med}_{n}(X)]P[X_{i}<\mbox{med}_{n}(X),~X_{j}<\mbox{med}_{n}(X)] \\
& + E[Z_{i}Z_{j}|X_{i}>\mbox{med}_{n}(X),~X_{j}>\mbox{med}_{n}(X)]P[X_{i}>\mbox{med}_{n}(X),~X_{j}>\mbox{med}_{n}(X)] \\
& + 2E[Z_{i}Z_{j}|X_{i}>\mbox{med}_{n}(X)>X_{j}]P[X_{i}>\mbox{med}_{n}(X)>X_{j}].
\end{align*}
The probabilities can be derived as follows 
\begin{align*}
P[X_{i}<\mbox{med}_{n}(X),~X_{j}<\mbox{med}_{n}(X)] & = \frac{{\frac{n}{2} \choose 2}}{{n \choose 2}} = \frac{n-2}{4(n-1)}, \\
P[X_{i}>\mbox{med}_{n}(X)>X_{j}] & = \frac{(\frac{n}{2})^2}{{n \choose 2}} = \frac{n}{4(n-1)}.
\end{align*}
By symmetry, we have
\begin{align*}
P[X_{i}>\mbox{med}_{n}(X),~X_{j}>\mbox{med}_{n}(X)] & = \frac{n-2}{4(n-1)}, \\
P[X_{i}<\mbox{med}_{n}(X)<X_{j}] & = \frac{n}{4(n-1)}.
\end{align*}
The conditional expectations can be derived as follows
\begin{align*}
E[Z_{i}Z_{j}|X_{i}<\mbox{med}_{n}(X),~X_{j}<\mbox{med}_{n}(X)] & = P(Z_{i}Z_{j} = 1) - P(Z_{i}Z_{j} = -1) = -\frac{1}{n-1}, \\
E[Z_{i}Z_{j}|X_{i}<\mbox{med}_{n}(X),~X_{j}<\mbox{med}_{n}(X)] & =  -\frac{1}{n-1}, \\
E[Z_{i}Z_{j}|X_{i}>\mbox{med}_{n}(X)>X_{j}] & =  \frac{1}{n-1}, \\
E[Z_{i}Z_{j}|X_{i}<\mbox{med}_{n}(X)<X_{j}] & =  \frac{1}{n-1}.
\end{align*}
Summarizing the results above, we have 
\begin{align*}
Cov(Z_{i},~Z_{j}) & = \frac{1}{(n-1)^2}, \\
V[Q_{n}(X, Y)] & = \frac{1}{n^2}[nV(Z_{i}) + n(n-1)Cov(Z_{i},~Z_{j})] \\
& = \frac{n}{n-1}.
\end{align*}

When $n$ is odd, we have $V(Z_{i}) = E(Z^{2}_{i}) = P(Z_{i}\neq 0) = (n-1)^2/n^2$. The covariance $Cov(Z_{i},~Z_{j}) = E(Z_{i}Z_{j})$ can be decomposed as
\begin{align*}
E(Z_{i}Z_{j}) = &~ 2E[Z_{i}Z_{j}|X_{i}<\mbox{med}_{n}(X),~X_{j}<\mbox{med}_{n}(X)]P[X_{i}<\mbox{med}_{n}(X),~X_{j}<\mbox{med}_{n}(X)] \\
& + 2E[Z_{i}Z_{j}|X_{i}>\mbox{med}_{n}(X)>X_{j}]P[X_{i}>\mbox{med}_{n}(X)>X_{j}].
\end{align*}
The probabilities can be derived as follows 
\begin{align*}
P[X_{i}<\mbox{med}_{n}(X),~X_{j}<\mbox{med}_{n}(X)] & = \frac{{\frac{n-1}{2} \choose 2}}{{n \choose 2}} = \frac{n-3}{4n}, \\
P[X_{i}>\mbox{med}_{n}(X)>X_{j}] & = \frac{(\frac{n-1}{2})^2}{{n \choose 2}} = \frac{n-1}{4n}.
\end{align*}
The conditional expectations are 
\begin{align*}
E[Z_{i}Z_{j}|X_{i}<\mbox{med}_{n}(X),~X_{j}<\mbox{med}_{n}(X)] & =  -\frac{1}{n}, \\
E[Z_{i}Z_{j}|X_{i}<\mbox{med}_{n}(X),~X_{j}<\mbox{med}_{n}(X)] & =  -\frac{1}{n}, \\
E[Z_{i}Z_{j}|X_{i}>\mbox{med}_{n}(X)>X_{j}] & =  \frac{1}{n}, \\
E[Z_{i}Z_{j}|X_{i}<\mbox{med}_{n}(X)<X_{j}] & =  \frac{1}{n}.
\end{align*}
Therefore we have
\begin{align*}
Cov(Z_{i},~Z_{j}) & = \frac{1}{n^2}, \\
V[Q_{n}(X, Y)] & = \frac{1}{n^2}[nV(Z_{i}) + n(n-1)Cov(Z_{i},~Z_{j})] \\
& = \frac{n-1}{n}.
\end{align*}

\subsection*{A.3. Proof of Lemma 4}
We first show the uncorrelatedness between $\xi_{n}(X, Y)$ and $\tau_{n}(X, Y)$. Ignoring the constants, we need show 
\begin{equation*}
Cov\left[\sum_{k=1}^{n-1}|R_{k+1}-R_{k}|,~\sum_{i<j}\mbox{sgn}(R_{i} - R_{j}) \right]= 0, 
\end{equation*}
or equivalently
\begin{equation*}
Cov\left[\sum_{k=1}^{n-1}R_{k+1}+\sum_{k=1}^{n-1}R_{k} - 2\sum_{k=1}^{n-1}\min(R_{k+1},R_{k}),~\sum_{i<j}\mbox{sgn}(R_{i} - R_{j}) \right]= 0.
\end{equation*}
We first derive $Cov[\sum_{k=1}^{n-1}R_{k+1}+\sum_{k=1}^{n-1}R_{k}, ~\sum_{i<j}\mbox{sgn}(R_{i} - R_{j})]$. For the ease of notation, for any $i<j$, define
\begin{align*}
A_{1} & =  Cov[R_{k},~ \mbox{sgn}(R_{i} - R_{j})],~\text{for~} k=i, \\
A_{2} & =  Cov[R_{k},~ \mbox{sgn}(R_{i} - R_{j})],~\text{for~}k=j, \\
A_{3} & =  Cov[R_{k},~ \mbox{sgn}(R_{i} - R_{j})],~\text{for~}k\neq i, j. 
\end{align*}
Under independence, $(R_{1},~...,~R_{n})$ is a random permutation of $(1,~...,~n)$. Furthermore, by symmetry, we have $A_{3} = 0$ and $A_{1} + A_{2} = 0$. To be specific, 
\begin{equation*}
A_{1} = \frac{(n+1)(n+5)}{6(n-1)} - \frac{2(n+2)}{n} - \frac{2}{n(n-1)}.
\end{equation*}
For $i=1$ and $j=n$,
\begin{equation*}
Cov\left[\sum_{k=1}^{n-1}R_{k+1}+\sum_{k=1}^{n-1}R_{k},~\mbox{sgn}(R_{i} - R_{j}) \right] = A_{1} + A_{2} = 0.
\end{equation*}
For $i=1$ and $j\neq n$,
\begin{equation*}
Cov\left[\sum_{k=1}^{n-1}R_{k+1}+\sum_{k=1}^{n-1}R_{k},~\mbox{sgn}(R_{i} - R_{j}) \right] = A_{2}.
\end{equation*}
For $i\neq 1$ and $j=n$,
\begin{equation*}
Cov\left[\sum_{k=1}^{n-1}R_{k+1}+\sum_{k=1}^{n-1}R_{k},~\mbox{sgn}(R_{i} - R_{j}) \right] = A_{1}.
\end{equation*}
For $i \neq 1, n$ and $j \neq 1, n$
\begin{equation*}
Cov\left[\sum_{k=1}^{n-1}R_{k+1}+\sum_{k=1}^{n-1}R_{k},~\mbox{sgn}(R_{i} - R_{j}) \right] = 0.
\end{equation*}
Summarizing the results above, we have
\begin{equation} \label{part1}
Cov\left[\sum_{k=1}^{n-1}R_{k+1}+\sum_{k=1}^{n-1}R_{k},~ \sum_{i<j} \mbox{sgn}(R_{i} - R_{j})\right] = 0.
\end{equation}
Similarly, for $Cov[\min(R_{k+1}, R_{k}),~\mbox{sgn}(R_{i}-R_{j})]$, we define
\begin{align*}
A_{4} & = Cov[\min(R_{k+1}, R_{k}),~\mbox{sgn}(R_{i}-R_{j})], \text{~for~} k\neq i, j,~ k+1 \neq i, j, \\
A_{5} & = Cov[\min(R_{k+1}, R_{k}),~\mbox{sgn}(R_{i}-R_{j})], \text{~for~} k=i,~ k+1= j, \\
A_{6} & = Cov[\min(R_{k+1}, R_{k}),~\mbox{sgn}(R_{i}-R_{j})], \text{~for~} k=i,~ k+1\neq j, \\
A_{7} & = Cov[\min(R_{k+1}, R_{k}),~\mbox{sgn}(R_{i}-R_{j})], \text{~for~} k+1=i,~ k\neq j, \\
A_{8} & = Cov[\min(R_{k+1}, R_{k}),~\mbox{sgn}(R_{i}-R_{j})], \text{~for~} k=j,~ k+1\neq i, \\
A_{9} & = Cov[\min(R_{k+1}, R_{k}),~\mbox{sgn}(R_{i}-R_{j})], \text{~for~} k+1=j,~ k\neq i,
\end{align*}
where by symmetry, we have $A_{4} = A_{5} = 0$, $A_{6} = A_{7}$, $A_{8}=A_{9}$ and $A_{6} + A_{8} = 0$. 

For $i=1$ and $j=n$, 
\begin{equation*}
Cov\left[\sum_{k=1}^{n-1}\min(R_{k+1}, R_{k}),~\mbox{sgn}(R_{i}-R_{j})\right] = A_{6} + A_{9} = 0. 
\end{equation*}

For $i=1$ and $j<n$, 
\begin{align*}
Cov\left[\sum_{k=1}^{n-1}\min(R_{k+1}, R_{k}),~\mbox{sgn}(R_{i}-R_{j})\right] & = A_{8}, \text{~for~} j=i+1 \\
Cov\left[\sum_{k=1}^{n-1}\min(R_{k+1}, R_{k}),~\mbox{sgn}(R_{i}-R_{j})\right] & = A_{6} + A_{7} + A_{8} = -A_{8}, \text{~for~} j>i+1. 
\end{align*}

For $1<i<j<n$, 
\begin{align*}
Cov\left[\sum_{k=1}^{n-1}\min(R_{k+1}, R_{k}),~\mbox{sgn}(R_{i}-R_{j})\right] & = A_{7}+A_{8} = 0, \text{~for~} j=i+1 \\
Cov\left[\sum_{k=1}^{n-1}\min(R_{k+1}, R_{k}),~\mbox{sgn}(R_{i}-R_{j})\right] & = A_{6} + A_{7} + A_{8} + A_{9}= 0, \text{~for~} j>i+1. 
\end{align*}

The covariance for $i>1$ and $j=n$ is same as the one for $i=1$ and $j<n$. Summarizing the results above, we have
\begin{equation} \label{part2}
Cov\left[\sum_{k=1}^{n-1}\min(R_{k+1}, R_{k}),~\sum_{i<j}\mbox{sgn}(R_{i}-R_{j})\right] = 0. 
\end{equation}
Therefore by Equations \ref{part1} and \ref{part2}, we have $Cov[\xi_{n}(X, Y),~\tau_{n}(X, Y)] =0$.

The proof for quadrant correlation is straightforward. Without loss of generality, we assume odd sample size 
\begin{equation*}
Q_{n}(X, Y) = \sum_{i = 1}^{n}\mbox{sgn}\left[\left(i-\frac{n+1}{2}\right)\left(R_{i}-\frac{n+1}{2} \right)\right].
\end{equation*}
Under independence, $(R_{1},~...,~R_{n})$ is a random permutation of $(1,~...,~n)$. By symmetry, we have 
\begin{align*}
Cov\left\{|R_{k+1}-R_{k}|,~\mbox{sgn}\left[\left(i-\frac{n+1}{2}\right)\left(R_{i}-\frac{n+1}{2} \right)\right] \right\} & = 0, \text{~for~} k = i, \\
Cov\left\{|R_{k+1}-R_{k}|,~\mbox{sgn}\left[\left(i-\frac{n+1}{2}\right)\left(R_{i}-\frac{n+1}{2} \right)\right] \right\} & = 0, \text{~for~} k+1 = i, \\
Cov\left\{|R_{k+1}-R_{k}|,~\mbox{sgn}\left[\left(i-\frac{n+1}{2}\right)\left(R_{i}-\frac{n+1}{2} \right)\right] \right\} & = 0, \text{~for~} k\neq i,~k+1\neq i.
\end{align*}
Therefore, $Cov[\xi_{n}(X, Y),~Q_{n}(X, Y)] =0$.

\subsection*{A.4. Calculation of $\mathrm{Cov}[|\tau_{n=3}(X,Y)|, \xi_{n=3}(X,Y)]$ }
When $n=3$ and $(X,~ Y)$ are independent, there are six equally probable permutations of ($R_{1}$, $R_{2}$, $R_{3}$). The table below summarizes the values of $\xi_{n=3}(X, Y)$, $\tau_{n=3}(X, Y)$, and $|\tau_{n=3}(X, Y)|$ for each permutation.
\begin{table}[!htbp]
\centering
\begin{tabular}{l r r r r}
\hline
($R_{1}$, $R_{2}$, $R_{3}$) & $\xi_{n=3}(X, Y)$ & $\tau_{n=3}(X, Y)$ & $|\tau_{n=3}(X, Y)|$\\ [0.5ex]
\hline
(1, 2, 3) & 1/4 & 1 & 1 \\
(1, 3, 2) & -1/8 &  1/3 & 1/3  \\
(2, 1, 3) & -1/8 & 1/3 & 1/3  \\
(2, 3, 1) & -1/8 & -1/3 & 1/3 \\
(3, 1, 2) & -1/8 & -1/3 & 1/3 \\
(3, 2, 1) & 1/4 & -1 & 1 \\ [1ex]
\hline 
\end{tabular}
\end{table}

Therefore we have $\mathrm{E}[\xi_{n=3}(X, Y)] = 0$ and 
\begin{align*}
\mathrm{Cov}[|\tau_{n=3}(X,Y)|, \xi_{n=3}(X,Y)] & = \mathrm{E}[\xi_{n=3}(X, Y)|\tau_{n=3}(X, Y)|] \\
& = \frac{1}{6}\left( 2\times \frac{1}{4}\times 1 - 4\times \frac{1}{8}\times\frac{1}{3}  \right) \\
& = \frac{1}{18}
\end{align*}

\subsection*{A.5. Proof of Theorem 2}
By the projection argument in H\'{a}jek (1968), $\tau_{n}(X,~Y)$ can be approximated by the following quantity \cite{Hajek}
$$\widetilde{\tau}_{n}(X,~Y) = \frac{8}{n^{2}(n-1)}\sum_{i=1}^{n}\left( i-\frac{n+1}{2} \right)\left( R(Y_{X(i)}) - \frac{n+1}{2} \right).$$
Under independence, Han et al. (2017) showed that $\widetilde{\tau}_{n}(X,~Y)$ and $\tau_{n}(X,~Y)$ are asymptotic equivalent (see \cite{Han17}, example 4, page 818). Recall that $S_{n}(X,~Y)$ can be rewritten as 
$$S_{n}(X,~Y) = \frac{12}{n(n-1)(n+1)}\sum_{i=1}^{n}\left( i-\frac{n+1}{2} \right)\left( R(Y_{X(i)}) - \frac{n+1}{2} \right).$$
Therefore under independence, $\tau_{n}(X,~Y)$ is asymptotically equivalent to $2S_{n}(X,~Y)/3$, and Theorem 1 also applies to Kendall's correlation. 

For quadrant correlation, similar to the proof of Lemma 2 in \cite{Zhang24}, we have 
\begin{equation*}
\frac{1}{\sqrt{n}}\sum_{i=1}^{n}\mbox{sgn}[(X_{i} - \mbox{med}_{n}(X))(Y_{i} - \mbox{med}_{n}(Y))] \stackrel{d}{\approx} \frac{1}{\sqrt{n}}\sum_{i=1}^{n}\mbox{sgn}\left[ \left( i-\frac{n}{2} \right) \left( V_{i} - \frac{1}{2} \right) \right].
\end{equation*}
where the right-hand side has expectation 0 and variance 1. Same as in Theorem 1, we will show the following quantity converges to a normal distribution for any $a,~b,~c\in \mathbb{R}$
\begin{align*}
J_{n} = \frac{1}{\sqrt{n}} \sum_{i=1}^{n} & \bigg[ a|F_{X}(X_{i})-F_{X}(X_{N_{Y}(i)})| + 2aF_{X}(X_{i})(1-F_{X}(X_{i}))  \\
& + b|F_{Y}(Y_{i})-F_{Y}(Y_{N_{X}(i)})| + 2bF_{Y}(Y_{i})(1-F_{Y}(Y_{i})) \\
& + c\cdot\mbox{sgn}\left( \frac{i}{n+1} - \frac{1}{2} \right)\mbox{sgn}\left( F_{X}(X_{i}) - \frac{1}{2} \right ) + \frac{2(a+b)}{3} \bigg].
\end{align*}

Same as in Theorem 1, it can be verified that $J_{n}(\mathcal{M})-J_{n}(\mathcal{M}^{i})-J_{n}(\mathcal{M}^{j})+J_{n}(\mathcal{M}^{ij}) = 0$ (see the definitions of $\mathcal{M},~\mathcal{M}^{i},~\mathcal{M}^{j},~\mathcal{M}^{ij}$ in A.1), i.e., $\mathcal{G}$ is a symmetric interaction rule. Using Theorem 2.5 of \cite{chatterjee.clt}, there exists a constant $C>0$, such that
$$\mathcal{D}(J_{n}) \leq \frac{C}{\sqrt{n}\sigma^2} + \frac{C}{2\sqrt{n}\sigma^3},$$
where $\sigma^{2} = \mbox{V}(J_{n})$, and $\mathcal{D}(J_{n})$ is the Wasserstein distance between $(J_{n}-E(J_{n}))/\sigma$ and $N(0 ,1)$. 

We now derive $\sigma^2$. Let $d_{i} = \mbox{sgn}[i/(n+1)-1/2]$, we have
\begin{align}
\sigma^{2}  = & V\left\{\frac{1}{\sqrt{n}}\sum_{i=1}^{n}\left[ a|U_{i+1}-U_{i}| + b|V_{i+1}-V_{i}| + 2aU_{i}(1-U_{i})  + 2bV_{i}(1-V_{i}) + c\cdot d_{i}\cdot \mbox{sgn}(U_{i}-\frac{1}{2}) \right]\right\} \nonumber \\
 = & V\left\{\frac{1}{\sqrt{n}}\sum_{i=1}^{n}[ a|U_{i+1}-U_{i}| + 2aU_{i}(1-U_{i})]\right\} + V\left\{\frac{1}{\sqrt{n}}\sum_{i=1}^{n}[ b|V_{i+1}-V_{i}| + 2bV_{i}(1-V_{i})]\right\} \label{pa6} \\
& + Cov\left\{\frac{1}{\sqrt{n}}\sum_{i=1}^{n}[ a|U_{i+1}-U_{i}| + 2aU_{i}(1-U_{i})],~ \frac{1}{\sqrt{n}}\sum_{i=1}^{n}[ b|V_{i+1}-V_{i}| + 2bV_{i}(1-V_{i})]\right\} \label{pa7}\\
& + Cov\left\{\frac{1}{\sqrt{n}}\sum_{i=1}^{n}[ a|U_{i+1}-U_{i}| + 2aU_{i}(1-U_{i})],~ \frac{1}{\sqrt{n}}\sum_{i=1}^{n}c\cdot d_{i}\cdot \mbox{sgn}(U_{i}-\frac{1}{2})\right\},\label{pa8} \\
& + Cov\left\{\frac{1}{\sqrt{n}}\sum_{i=1}^{n}[ b|V_{i+1}-V_{i}| + 2bV_{i}(1-V_{i})],~ \frac{1}{\sqrt{n}}\sum_{i=1}^{n}c\cdot d_{i}\cdot \mbox{sgn}(U_{i}-\frac{1}{2})\right\},\label{pa9} \\
& + V\left(\frac{1}{\sqrt{n}}\sum_{i=1}^{n}c\cdot d_{i}\cdot \mbox{sgn}(U_{i}-\frac{1}{2})\right). \label{pa10}
\end{align}
From Equation (14) in \cite{angus}, term (\ref{pa6}) is $2(a^2+b^2)/45 + O(1/n)$. From Lemma 3 of \cite{Zhang23}, term (\ref{pa7}) is $O(1/n)$. Using the facts that for two independent variables $(U_{1},~U_{2})\sim \mbox{Uniform}(0,~ 1)$, $Cov[\mbox{sgn}(U_{1}-1/2),~U_{1}] = 1/4$, $Cov[\mbox{sgn}(U_{1}-1/2),~U^2_{1}] = 1/4$ and $Cov[|U_{1}-U_{2}|,~\mbox{sgn}(U_{1}-1/2)] = 0$, it can be shown that for term (\ref{pa8}) 
$$Cov\left\{\frac{1}{\sqrt{n}}\sum_{i=1}^{n}[ a|U_{i+1}-U_{i}| + 2aU_{i}(1-U_{i})],~ \frac{1}{\sqrt{n}}\sum_{i=1}^{n}c\cdot d_{i}\cdot \mbox{sgn}(U_{i}-\frac{1}{2})\right\}  = 0.$$
For term (\ref{pa10}), we have
$$V\left(\frac{1}{\sqrt{n}}\sum_{i=1}^{n}c\cdot d_{i}\cdot \mbox{sgn}(U_{i}-\frac{1}{2})\right) = c^2.$$
Term (\ref{pa9}) is bounded by 
$$ \frac{1}{2}V\left(\frac{1}{\sqrt{n}}\sum_{i=1}^{n}c\cdot d_{i}\cdot \mbox{sgn}(U_{i}-\frac{1}{2})\right) + \frac{1}{2} V\left\{\frac{1}{\sqrt{n}}\sum_{i=1}^{n}[ b|V_{i+1}-V_{i}| + 2bV_{i}(1-V_{i})]\right\} = \frac{b^2}{45}+ \frac{c^2}{2} + O\left( \frac{1}{n} \right).$$ 
Therefore we have 
$$\frac{2a^2}{45}+ \frac{b^2}{45}+ \frac{c^2}{2} + O\left( \frac{1}{n} \right) \leq \sigma^{2} \leq \frac{2a^2}{45} + \frac{b^2}{15} + \frac{3c^2}{2} +O\left( \frac{1}{n} \right), $$
and the Wasserstein distance between $(J_{n}-E(J_{n}))/\sigma$ and $N(0 ,1)$ converges to 0. This completes the proof.

\subsection*{A.6. Binary expansion by Chatterjee (2022)}
Let $x=(x_{1}, ..., x_{d})$ be any vector in $\mathbb{R}^{d}$. The binary expansion mapping $\eta(x)$ is essentially an encoding of the $d$-tuple of $(x_{1}, ..., x_{d})$. Starting from the absolute values $(|x_{1}|, ..., |x_{d}|)$, let 
$$a_{i1}a_{i2}...a_{ik_{i}}\textbf{.}~b_{i1}b_{i2}...$$
be the binary expansion of $|x_{i}|$. By filling in extra 0's at the beginning, we can make $k_{1} = ... = k_{d} = k$. The $d$-tuple $(|x_{1}|, ..., |x_{d}|)$ can then be mapped to
$$a_{11}a_{21}...a_{d1}a_{12}...a_{d2}...a_{1k}...a_{dk}\textbf{.}~b_{11}b_{21}...b_{d1}b_{12}...b_{d2}... .$$
Next, we can encode the signs of $x_{i}$'s by $1c_{1}...c_{d}$, where $c_{i}=1$ if $x_{i}\geq 0$ and 0 if $x_{i}<0$ (the 1 in front ensures there is no leading 0). Lastly, the signed $d$-tuple $x$ is mapped to
$$1c_{1}...c_{d}a_{11}a_{21}...a_{d1}a_{12}...a_{d2}...a_{1k}...a_{dk}\textbf{.}~b_{11}b_{21}...b_{d1}b_{12}...b_{d2}... ,$$
which is a Borel isomorphism as illustrated by \cite{chatterjee.survey}.

\newpage
\section*{Tables and Figures}
\begin{table}[H]
\centering
\title{Table 1: Empirical power for univariate $X$ and $Y$}\vspace{2mm}
\rowcolors{1}{}{lightgray}
\begin{tabular}{p{19mm}p{15mm}p{17mm}p{17mm}p{17mm}p{17mm}p{17mm}p{17mm}p{10mm}}
  \hline  \hline
setting & n & $\widetilde{I}^{S}_{n}$ & $\widetilde{I}^{\tau}_{n}$ & $\widetilde{I}^{Q}_{n}$ & $\xi^{sym}_{n}$ & $D$ & $R$ & $\tau^*$  \vspace{1.5mm} \\
  \hline
Linear &   20 & 0.500 & 0.562 & 0.375 & 0.247 & 0.521 & 0.561 & 0.564 \vspace{1.5mm}\\ 
 &   40 & 0.863 & 0.876 & 0.563 & 0.440 & 0.871 & 0.885 & 0.883 \vspace{1.5mm}\\ 
 &   60 & 0.971 & 0.973 & 0.809 & 0.608 & 0.974 & 0.977 & 0.977\vspace{1.5mm} \\ 
 &   80 & 0.996 & 0.997 & 0.920 & 0.708 & 0.997 & 0.997 & 0.997\vspace{1.5mm} \\ 
&  100 & 0.999 & 0.999 & 0.951 & 0.796 & 0.999 & 0.999 & 0.999\vspace{1.5mm} \\ 
\hline
Quadratic &   20 & 0.347 & 0.364 & 0.354 & 0.456 & 0.282 & 0.213 & 0.252 \vspace{1.5mm}\\ 
 &   40 & 0.747 & 0.750 & 0.742 & 0.817 & 0.772 & 0.739 & 0.762\vspace{1.5mm} \\ 
 &   60 & 0.915 & 0.915 & 0.915 & 0.948 & 0.971 & 0.968 & 0.972 \vspace{1.5mm}\\ 
&   80 & 0.966 & 0.967 & 0.967 & 0.981 & 0.998 & 0.997 & 0.998\vspace{1.5mm} \\ 
  &  100 & 0.992 & 0.992 & 0.991 & 0.996 & 1.000 & 1.000 & 1.000 \vspace{1.5mm}\\ 
 \hline
 Stepwise &   20 & 0.446 & 0.508 & 0.348 & 0.217 & 0.477 & 0.511 & 0.517\vspace{1.5mm} \\ 
 &   40 & 0.809 & 0.827 & 0.527 & 0.397 & 0.830 & 0.851 & 0.848\vspace{1.5mm} \\ 
 &   60 & 0.951 & 0.956 & 0.776 & 0.551 & 0.957 & 0.963 & 0.963\vspace{1.5mm} \\ 
&   80 & 0.990 & 0.990 & 0.894 & 0.660 & 0.992 & 0.992 & 0.993\vspace{1.5mm} \\ 
&  100 & 0.998 & 0.998 & 0.943 & 0.750 & 0.998 & 0.998 & 0.998 \vspace{1.5mm}\\ 
\hline
Sinusoid &   20 & 0.184 & 0.198 & 0.202 & 0.283 & 0.079 & 0.075 & 0.080\vspace{1.5mm} \\ 
 &   40 & 0.677 & 0.678 & 0.678 & 0.765 & 0.139 & 0.116 & 0.118\vspace{1.5mm} \\ 
 &   60 & 0.888 & 0.888 & 0.890 & 0.931 & 0.206 & 0.154 & 0.172 \vspace{1.5mm}\\ 
 &   80 & 0.962 & 0.962 & 0.963 & 0.979 & 0.358 & 0.234 & 0.272\vspace{1.5mm} \\ 
 &  100 & 0.986 & 0.986 & 0.986 & 0.992 & 0.534 & 0.375 & 0.422 \\ 
  \hline  \hline
\end{tabular}
\caption*{Presented in the table above are the empirical power of seven independence tests: Chatterjee-Spearman ($\widetilde{I}^{S}_{n}$), Chatterjee-Kendall ($\widetilde{I}^{\tau}_{n}$), Chatterjee-quadrant ($\widetilde{I}^{Q}_{n}$), symmetrized Chatterjee ($\xi^{sym}_{n}$), Hoeffding's $D$, Blum-Kiefer-Rosenblatt's $R$, and Bergsma-Dassios' $\tau^*$.}
\end{table}

\newpage
\begin{table}[H]
\centering
\title{Table 2: Empirical size for univariate $X$ and $Y$}\vspace{2mm}
\rowcolors{1}{}{lightgray}
\begin{tabular}{p{22mm}p{19mm}p{19mm}p{19mm}p{19mm}p{19mm}p{19mm}p{19mm}p{10mm}}
  \hline  \hline
n & $\widetilde{I}^{S}_{n}$ & $\widetilde{I}^{\tau}_{n}$ & $\widetilde{I}^{Q}_{n}$ & $\xi^{sym}_{n}$ & $D$ & $R$ & $\tau^*$  \vspace{1.5mm} \\
  \hline
 20 & 0.038 & 0.051 & 0.048 & 0.037 & 0.048 & 0.049 & 0.052 \vspace{1.5mm}\\ 
    40 &  0.045 & 0.050 & 0.046 & 0.043 & 0.051 & 0.053 & 0.050 \vspace{1.5mm}\\ 
    60 & 0.046 & 0.050 & 0.046 & 0.045 & 0.050 & 0.049 & 0.048 \vspace{1.5mm} \\ 
    80 & 0.047 & 0.049 & 0.051 & 0.047 & 0.049 & 0.050 & 0.051 \vspace{1.5mm} \\ 
  100 & 0.047 & 0.050 & 0.049 & 0.048 & 0.048 & 0.048 & 0.047 \vspace{1.5mm} \\   \hline  \hline
\end{tabular}
\caption*{Presented in the table above are the empirical size of seven independence tests: Chatterjee-Spearman ($\widetilde{I}^{S}_{n}$), Chatterjee-Kendall ($\widetilde{I}^{\tau}_{n}$), Chatterjee-quadrant ($\widetilde{I}^{Q}_{n}$), symmetrized Chatterjee ($\xi^{sym}_{n}$), Hoeffding's $D$, Blum-Kiefer-Rosenblatt's $R$, and Bergsma-Dassios' $\tau^*$.}
\end{table}

\newpage
\begin{table}[H]
\centering
\title{Table 3: Empirical power for multivariate $\mathbf{X}$ and $\mathbf{Y}$ based on Grothe et al.'s formulas}\vspace{2mm}
\rowcolors{1}{}{lightgray}
\begin{tabular}{p{19mm}p{17mm}p{24mm}p{24mm}p{24mm}p{24mm}p{24mm}}
  \hline  \hline
setting & n & $\xi^{sym}_{n}$ & $S_{n}$ & $\tau_{n}$ & $\widetilde{I}^{S}_{n}$ &  $\widetilde{I}^{\tau}_{n}$  \vspace{1.5mm} \\
  \hline
1 & 20 & 0.091 & 0.293 & 0.400 & 0.251 & 0.375 \vspace{1.5mm}\\ 
 &   40 &  0.135 & 0.435 & 0.659 & 0.427 & 0.648 \vspace{1.5mm}\\ 
 &   60 &  0.170 & 0.655 & 0.862 & 0.635 & 0.831 \vspace{1.5mm} \\ 
 &   80 & 0.182 & 0.733 & 0.860 & 0.708 & 0.872 \vspace{1.5mm} \\ 
   \hline
2 & 20 &  0.057 & 0.244 & 0.445 & 0.240 & 0.415 \vspace{1.5mm}\\ 
 &   40 & 0.079 & 0.361 & 0.508 & 0.350 & 0.487 \vspace{1.5mm}\\ 
 &   60 &  0.111 & 0.531 & 0.730 & 0.524 & 0.719  \vspace{1.5mm} \\ 
 &   80 & 0.139 & 0.621 & 0.810 & 0.591  & 0.799  \vspace{1.5mm} \\ 
   \hline
3 & 20 & 0.090  & 0.342  & 0.509  & 0.330  & 0.495  \vspace{1.5mm}\\ 
 &   40 &  0.159  & 0.513  &  0.750 &  0.501 & 0.733  \vspace{1.5mm}\\ 
 &   60 & 0.198  & 0.650  & 0.838  & 0.643  & 0.822  \vspace{1.5mm} \\ 
 &   80 & 0.237 & 0.821 & 0.941  & 0.803 &  0.934 \vspace{1.5mm} \\ 
\hline
4 & 20 & 0.451 & 0.075 & 0.080 & 0.351 & 0.368 \vspace{1.5mm}\\ 
 &   40 &  0.692 & 0.137 & 0.083 & 0.475 & 0.454 \vspace{1.5mm}\\ 
 &   60 & 0.873 & 0.165 & 0.090 & 0.825 & 0.805 \vspace{1.5mm} \\ 
 &   80 & 0.983 & 0.178 & 0.099 & 0.946 & 0.939 \vspace{1.5mm} \\ 
  \hline
 5 & 20 &  0.512 & 0.110 & 0.115  & 0.290  & 0.325   \vspace{1.5mm}\\ 
 &   40 & 0.785  & 0.113  & 0.116 &  0.564 & 0.565  \vspace{1.5mm}\\ 
 &   60 &  0.933 & 0.112  & 0.120  &  0.741 & 0.754  \vspace{1.5mm} \\ 
 &   80 & 0.976  & 0.126  &  0.131 & 0.810  & 0.812  \vspace{1.5mm} \\ 
\hline
6 & 20 & 0.105  & 0.052  &  0.056 & 0.061  & 0.075  \vspace{1.5mm}\\ 
 &   40 &  0.216  & 0.053  &  0.060 & 0.082  & 0.114  \vspace{1.5mm}\\ 
 &   60 & 0.341 & 0.056  & 0.098  & 0.159  &  0.178 \vspace{1.5mm} \\ 
 &   80 & 0.553  & 0.058 & 0.110 & 0.289  & 0.337  \vspace{1.5mm} \\ 
  \hline  \hline
\end{tabular}
\caption*{Presented in the table above are the empirical power of five independence tests (permutation-based): symmetrized Chatterjee ($\xi^{sym}_{n}$), Spearman ($S_{n}$), Kendall ($\tau_{n}$), Chatterjee-Spearman ($\widetilde{I}^{S}_{n}$), and Chatterjee-Kendall ($\widetilde{I}^{\tau}_{n}$), where $S_{n}$ and $\tau_{n}$ are based on Grothe et al.'s formulas \cite{skvector}.}
\end{table}

\newpage
\begin{table}[H]
\centering
\title{Table 4: Empirical power for multivariate $\mathbf{X}$ and $\mathbf{Y}$ based on Borel isomorphism}\vspace{2mm}
\rowcolors{1}{}{lightgray}
\begin{tabular}{p{15mm}p{13mm}p{26mm}p{26mm}p{26mm}p{26mm}p{26mm}}
  \hline  \hline
setting & n & $\xi^{sym}_{n}$ & $S_{n}$ & $\tau_{n}$ & $\widetilde{I}^{S}_{n}$ &  $\widetilde{I}^{\tau}_{n}$  \vspace{1.5mm} \\
  \hline
1 & 20 & 0.087 (0.091) & 0.165 (0.153) & 0.190 (0.162) & 0.111 (0.146) & 0.140 (0.149) \vspace{1.5mm}\\ 
 &   40 &  0.097 (0.135) & 0.272 (0.261) & 0.301 (0.288) & 0.212 (0.231) & 0.233 (0.256) \vspace{1.5mm}\\ 
 &   60 & 0.139 (0.170) & 0.389 (0.380) & 0.412 (0.373) & 0.310 (0.346) & 0.339 (0.327) \vspace{1.5mm} \\ 
 &   80 & 0.154 (0.182) & 0.509 (0.539) & 0.527 (0.530) & 0.409 (0.525) & 0.429 (0.518) \vspace{1.5mm} \\ 
  \hline
2 & 20 & 0.048 (0.057) & 0.052 (0.055) & 0.066 (0.061) & 0.050 (0.055) & 0.059 (0.058)  \vspace{1.5mm}\\ 
 &   40 & 0.054 (0.079) & 0.056 (0.057) & 0.072 (0.063) & 0.054 (0.059) & 0.061 (0.066) \vspace{1.5mm}\\ 
 &   60 & 0.058 (0.111) & 0.060 (0.060) & 0.077 (0.069) & 0.060 (0.064) & 0.068 (0.076) \vspace{1.5mm} \\ 
 &   80 & 0.111 (0.139) & 0.068 (0.059) & 0.080 (0.081) & 0.104 (0.085) & 0.109 (0.097) \vspace{1.5mm} \\ 
  \hline
3 & 20 & 0.066 (0.090) & 0.152 (0.147) & 0.187 (0.180) & 0.109 (0.145) & 0.128 (0.169) \vspace{1.5mm}\\ 
 &   40 & 0.134 (0.159) & 0.252 (0.268) & 0.273 (0.270) & 0.202 (0.263) & 0.228 (0.267) \vspace{1.5mm}\\ 
 &   60 & 0.182 (0.198) & 0.352 (0.405) & 0.384 (0.419) & 0.309 (0.355) & 0.334 (0.408) \vspace{1.5mm} \\ 
 &   80 & 0.218 (0.237) & 0.512 (0.566) & 0.537 (0.578) & 0.450 (0.476) & 0.461 (0.501)  \vspace{1.5mm} \\ 
\hline
4 & 20 & 0.305 (0.451) & 0.569 (0.549) & 0.643 (0.605) & 0.485 (0.534) & 0.554 (0.583)  \vspace{1.5mm}\\ 
 &   40 & 0.606 (0.692) & 0.868 (0.858) & 0.901 (0.873) & 0.833 (0.827) & 0.865 (0.844) \vspace{1.5mm}\\ 
 &   60 & 0.805 (0.853) & 0.963 (0.964) & 0.969 (0.967) &  0.961 (0.951) & 0.968(0.952) \vspace{1.5mm} \\ 
 &   80 & 0.916 (0.983) & 0.994 (0.998) & 0.995 (1.000) & 0.994 (0.997) & 0.994 (1.000) \vspace{1.5mm} \\ 
 \hline
  5 & 20 & 0.462 (0.512) & 0.778 (0.744) & 0.801 (0.723) & 0.676 (0.740) & 0.748 (0.711)  \vspace{1.5mm}\\ 
 &   40 &  0.734 (0.785) & 0.978 (0.981) & 0.982 (0.976) & 0.966 (0.972) & 0.967 (0.971)  \vspace{1.5mm}\\ 
 &   60 & 0.918 (0.933) & 0.998 (0.999) & 0.998 (1.000) & 0.997 (0.999)  & 0.997 (0.999) \vspace{1.5mm} \\ 
 &   80 & 0.956 (0.976) & 1.000 (0.999) & 1.000 (1.000) & 1.000 (0.999) & 1.000 (0.999)  \vspace{1.5mm} \\ 
\hline
6 & 20 & 0.094 (0.105) & 0.086 (0.059) & 0.100 (0.057) & 0.091 (0.090)  & 0.098 (0.088)   \vspace{1.5mm}\\ 
 &   40 & 0.209 (0.216) & 0.124 (0.062) & 0.136 (0.064) & 0.180 (0.105) & 0.186 (0.105)  \vspace{1.5mm}\\ 
 &   60 & 0.322 (0.341) & 0.136 (0.072) & 0.144 (0.069) & 0.268 (0.149) & 0.272 (0.148)  \vspace{1.5mm} \\ 
 &   80 & 0.536 (0.553) & 0.214 (0.080) & 0.218 (0.081) & 0.508 (0.251) & 0.510 (0.253)  \vspace{1.5mm} \\ 
  \hline  \hline
\end{tabular}
\caption*{Presented in the table above are the empirical power of five independence tests based on analytical and permutation p-values (permutation tests are shown in parentheses): symmetrized Chatterjee ($\xi^{sym}_{n}$), Spearman ($S_{n}$), Kendall ($\tau_{n}$), Chatterjee-Spearman ($\widetilde{I}^{S}_{n}$), and Chatterjee-Kendall ($\widetilde{I}^{\tau}_{n}$), where $S_{n}$ and $\tau_{n}$ are based on Borel isomorphism \cite{chatterjee.survey}.}
\end{table}

\newpage
\begin{table}[H]
\centering
\title{Table 5: Empirical size for multivariate $\mathbf{X}$ and $\mathbf{Y}$ based on Borel isomorphism}\vspace{2mm}
\rowcolors{1}{}{lightgray}
\begin{tabular}{p{15mm}p{26mm}p{26mm}p{26mm}p{26mm}p{26mm}p{26mm}}
  \hline  \hline
 n & $\xi^{sym}_{n}$ & $S_{n}$ & $\tau_{n}$ & $\widetilde{I}^{S}_{n}$ &  $\widetilde{I}^{\tau}_{n}$  \vspace{1.5mm} \\
  \hline
20 & 0.034 (0.051) & 0.059 (0.048) & 0.078 (0.051) &  0.036 (0.052) & 0.055 (0.051) \vspace{1.5mm}\\ 
40 & 0.047 (0.047) & 0.052 (0.051) & 0.061 (0.051) &  0.045 (0.053) & 0.051 (0.050) \vspace{1.5mm}\\ 
60 & 0.048 (0.049) & 0.049 (0.052) & 0.056 (0.048) &  0.048 (0.051) & 0.050 (0.048) \vspace{1.5mm} \\ 
80 & 0.048 (0.052) & 0.050 (0.050) & 0.055 (0.048) &  0.049 (0.049) & 0.051 (0.052) \vspace{1.5mm} \\ 
  \hline  \hline
\end{tabular}
\caption*{Presented in the table above are the empirical size of five independence tests based on analytical and permutation p-values (permutation tests are shown in parentheses): symmetrized Chatterjee ($\xi^{sym}_{n}$), Spearman ($S_{n}$), Kendall ($\tau_{n}$), Chatterjee-Spearman ($\widetilde{I}^{S}_{n}$), and Chatterjee-Kendall ($\widetilde{I}^{\tau}_{n}$), where $S_{n}$ and $\tau_{n}$ are based on Borel isomorphism \cite{chatterjee.survey}.}
\end{table}

\newpage
\begin{figure}[!htbp]
\begin{center}
\includegraphics[scale=0.7]{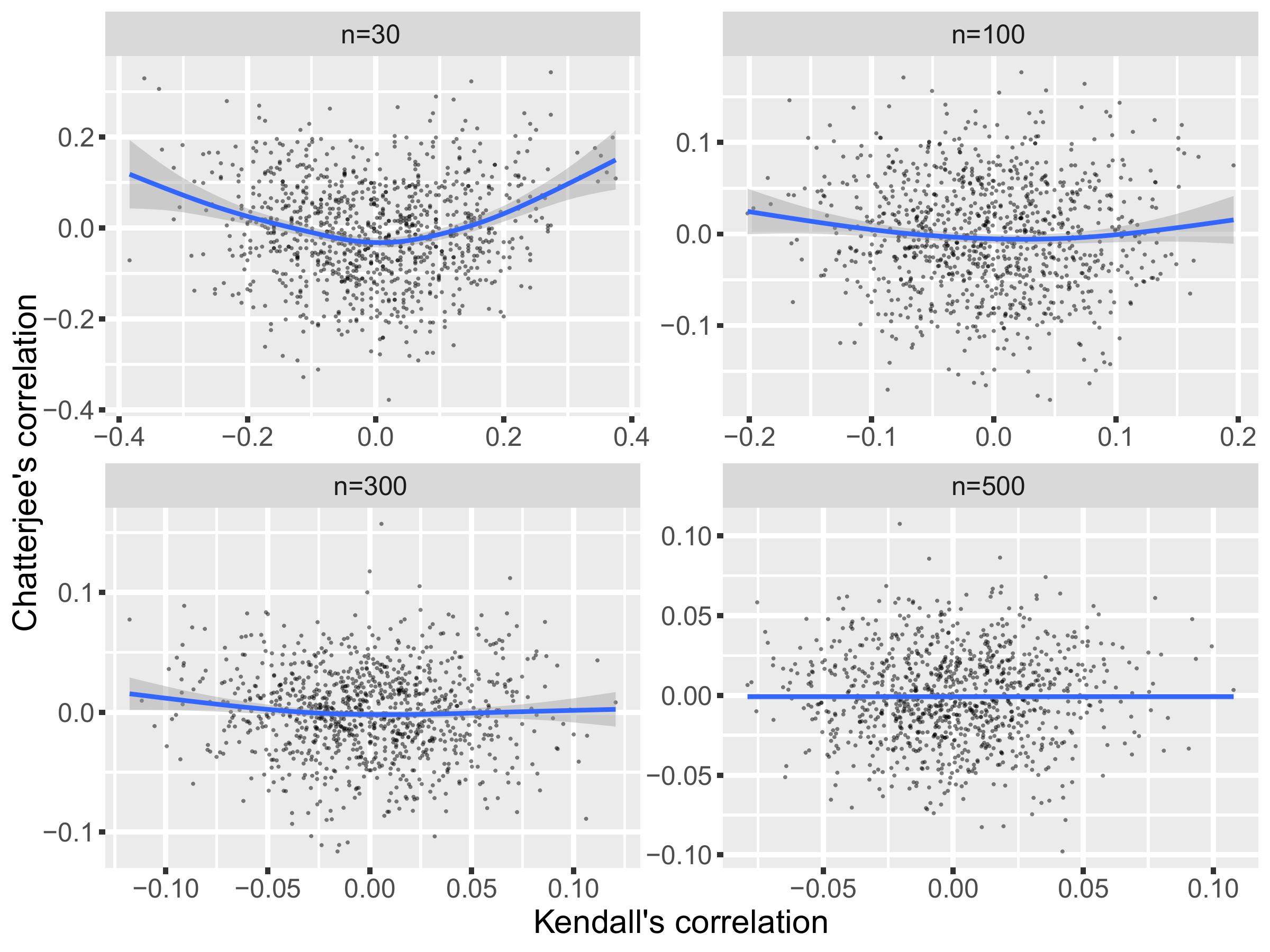}
\end{center}
\caption{Scatterplots of $\tau_{n}(X, Y)$ and $\xi_{n}(X, Y)$ under $n = 30, 100, 300, 500$. The blue line is the LOESS smoothing function, and the gray area is the 95\% confidence interval. The data were generated using $X \sim \mbox{Uniform}[0, 1]$, $Y\sim N(0,1)$.
}
\end{figure}

\newpage
\begin{figure}[!htbp]
\begin{center}
\includegraphics[scale=0.7]{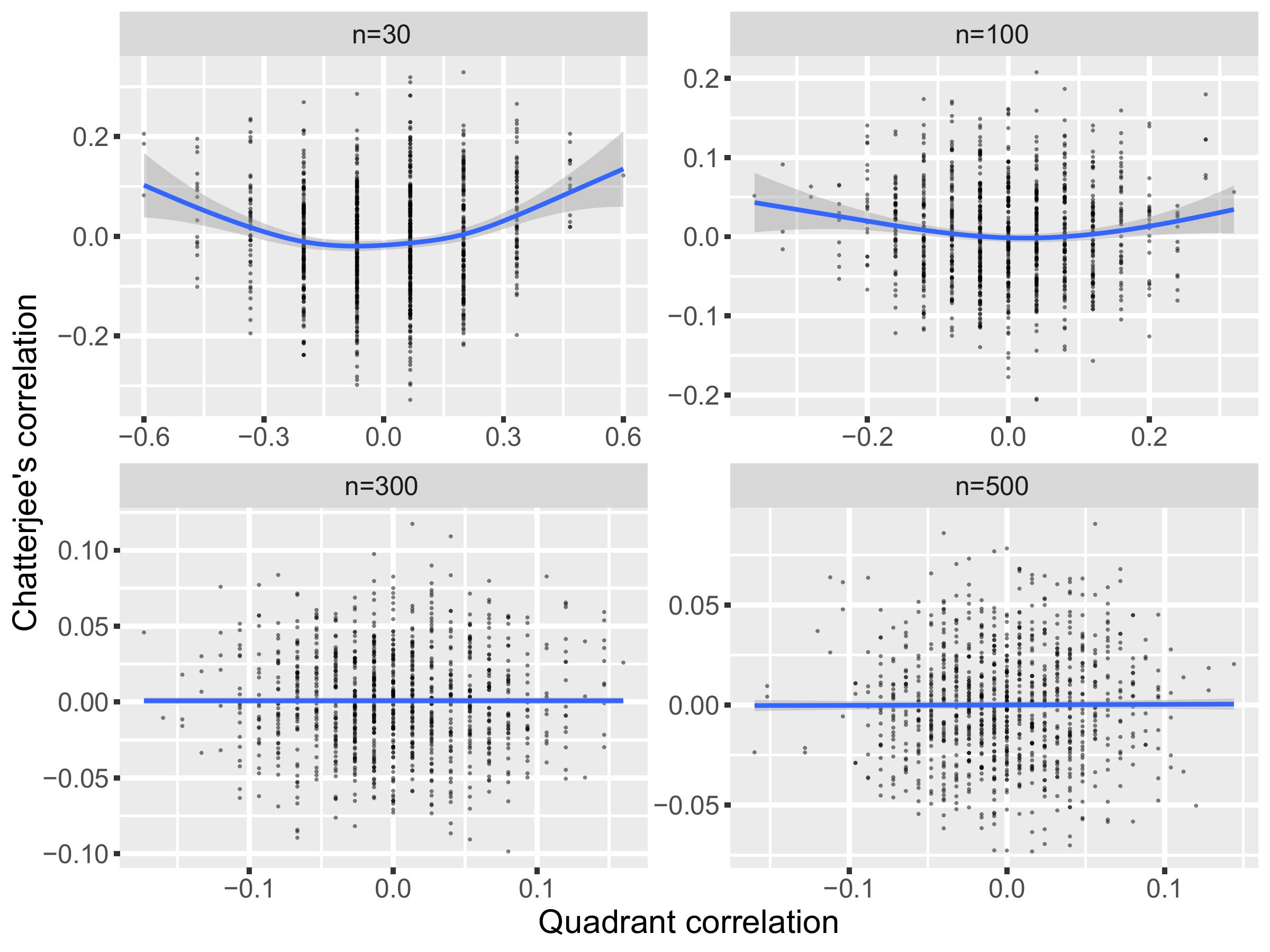}
\end{center}
\caption{Scatterplots of $Q_{n}(X, Y)$ and $\xi_{n}(X, Y)$ under $n = 30, 100, 300, 500$. The blue line is the LOESS smoothing function, and the gray area is the 95\% confidence interval. The data were generated using $X \sim \mbox{Uniform}[0, 1]$, $Y\sim N(0,1)$.
}
\end{figure}

\newpage
\begin{figure}[!htbp]
\begin{center}
\includegraphics[scale=0.7]{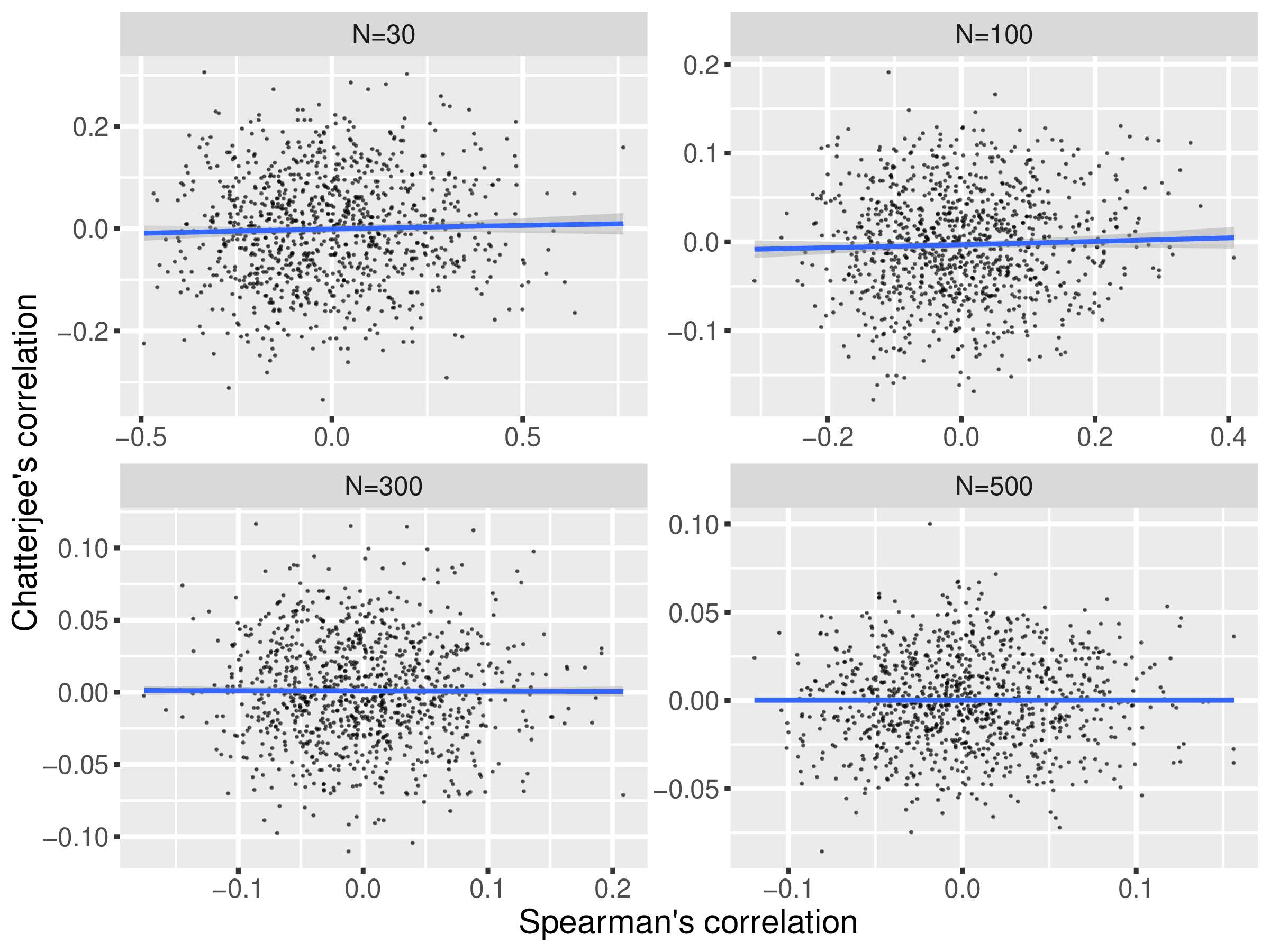}
\end{center}
\caption{Scatterplots of $S_{n}(\pmb{X}, \pmb{Y})$ and $\xi_{n}(\pmb{X}, \pmb{Y})$ under $n = 30, 100, 300, 500$, where $S_{n}(\pmb{X}, \pmb{Y})$ is based on Grothe et al.'s formulas \cite{skvector}. The blue line is the LOESS smoothing function, and the gray area is the 95\% confidence interval. The data were generated using $\mathbf{X} = (X_{[1]},~X_{[2]},~X_{[3]}),~\mathbf{Y} = (Y_{[1]},~Y_{[2]},~Y_{[3]})$, $X_{[i]} \sim \mbox{Uniform}[0, 1]$, $Y_{[i]}\sim N(0,1)$, $i = 1,~2,~3.$
}
\end{figure}

\newpage
\begin{figure}[!htbp]
\begin{center}
\includegraphics[scale=0.7]{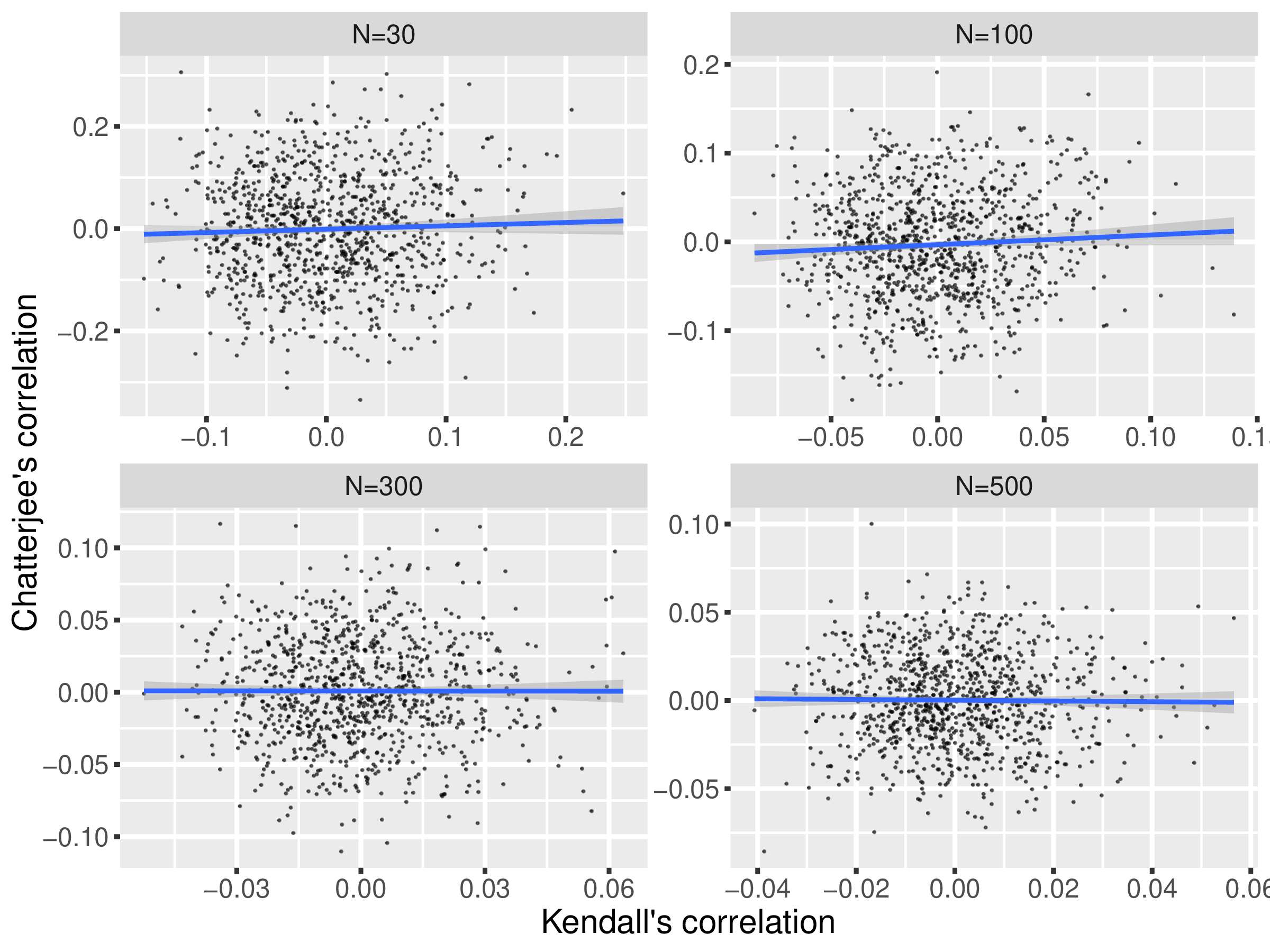}
\end{center}
\caption{Scatterplots of $\tau_{n}(\pmb{X}, \pmb{Y})$ and $\xi_{n}(\pmb{X}, \pmb{Y})$ under $n = 30, 100, 300, 500$, where $\tau_{n}(\pmb{X}, \pmb{Y})$ is based on Grothe et al.'s formulas \cite{skvector}. The blue line is the LOESS smoothing function, and the gray area is the 95\% confidence interval. The data were generated using $\mathbf{X} = (X_{[1]},~X_{[2]},~X_{[3]}),~\mathbf{Y} = (Y_{[1]},~Y_{[2]},~Y_{[3]})$, $X_{[i]} \sim \mbox{Uniform}[0, 1]$, $Y_{[i]}\sim N(0,1)$, $i = 1,~2,~3.$
}
\end{figure}

\newpage
\begin{figure}[!htbp]
\begin{center}
\includegraphics[scale=0.7]{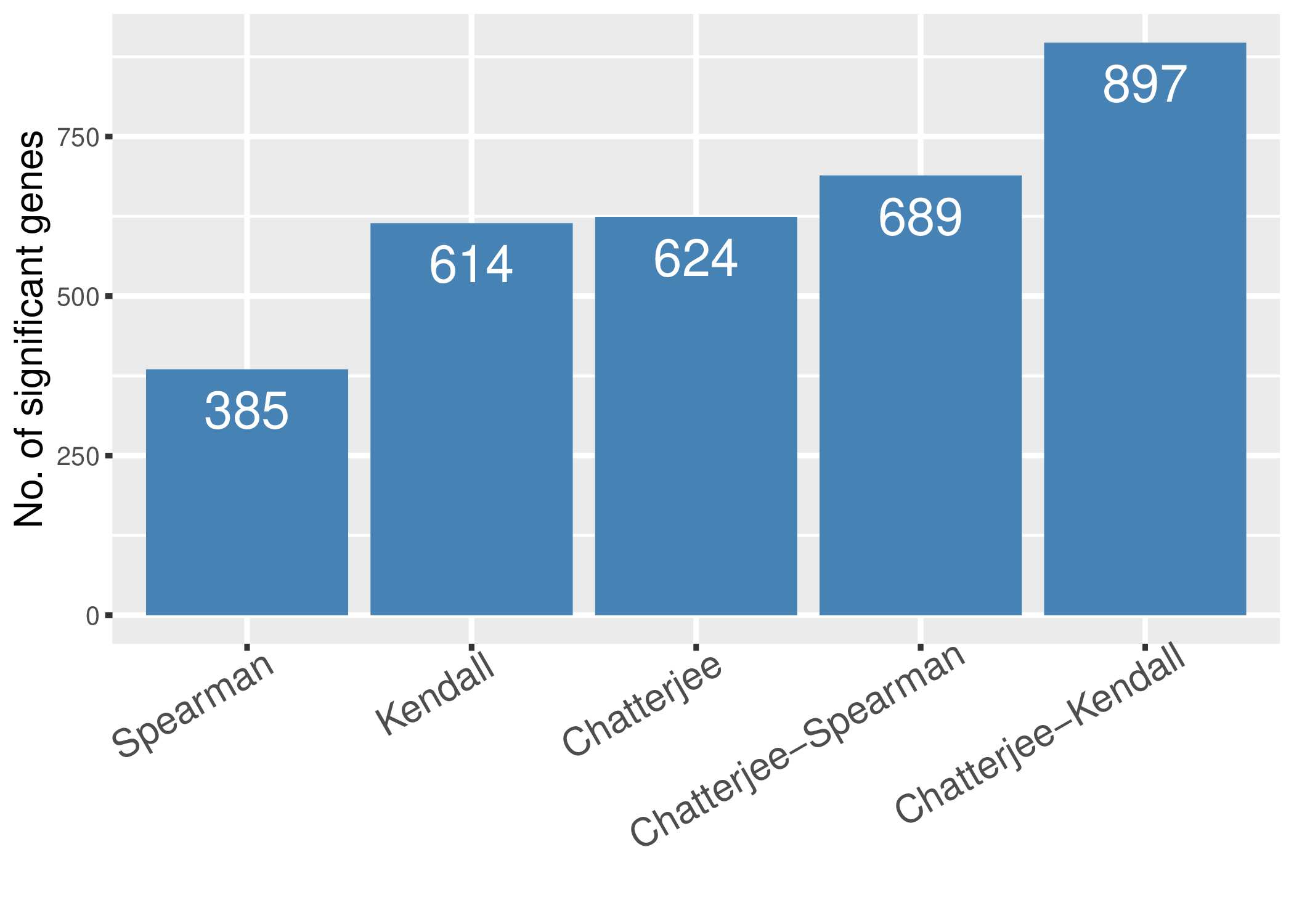}
\end{center}
\caption{Number of significant genes identified by five tests.
}
\end{figure}

\newpage
\begin{figure}[!htbp]
\begin{center}
\includegraphics[scale=0.65]{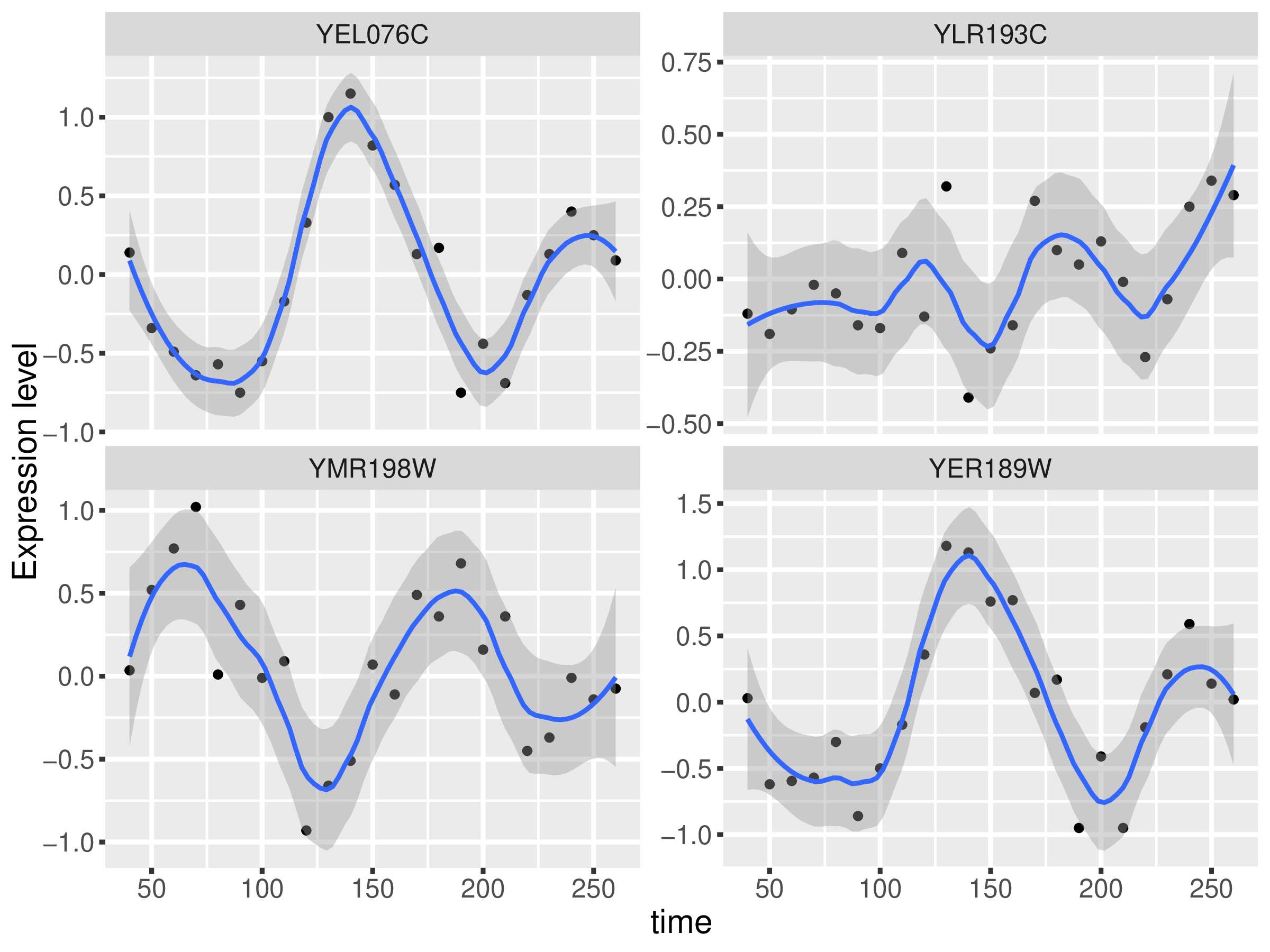}
\end{center}
\caption{A random sample of 4 genes selected by the Chatterjee-Kendall test but missed by Kendall's test. The blue line represents the LOESS smoothing function, and the gray area shows the 95\% confidence interval.
}
\end{figure}

\newpage
\begin{figure}[!htbp]
\begin{center}
\includegraphics[scale=0.65]{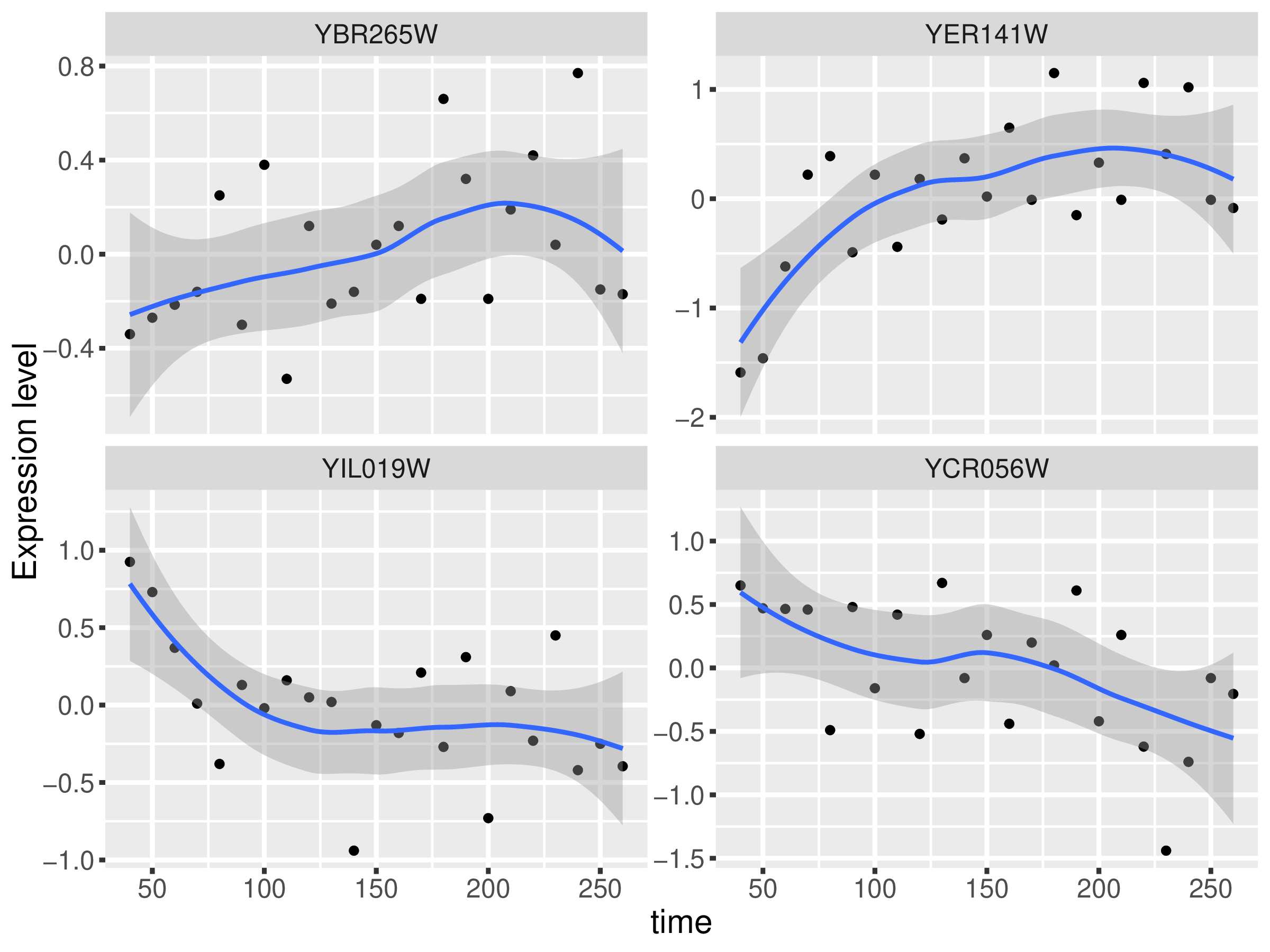}
\end{center}
\caption{A random sample of 4 genes selected by the Chatterjee-Kendall test but missed by Chatterjee's test. The blue line represents the LOESS smoothing function, and the gray area shows the 95\% confidence interval.
}
\end{figure}

\newpage
\begin{figure}[!htbp]
\begin{center}
\includegraphics[scale=0.65]{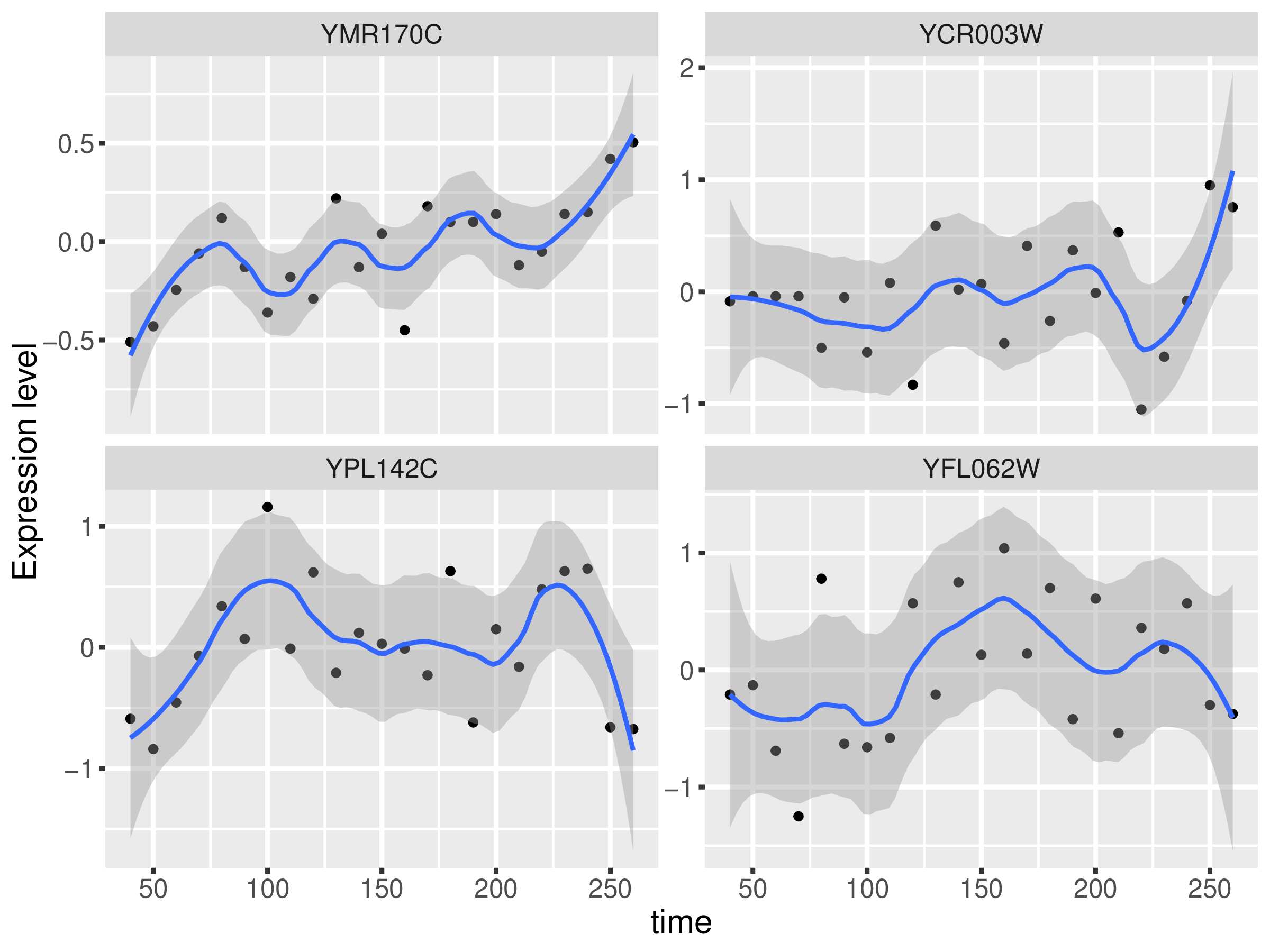}
\end{center}
\caption{A random sample of 4 genes selected by the Chatterjee-Kendall test but missed by the Chatterjee-Spearman test. The blue line represents the LOESS smoothing function, and the gray area shows the 95\% confidence interval.
}
\end{figure}


\begin{thebibliography}{9}
\bibitem{crouxdehon} Croux, C. \& Dehon, C. (2010). Influence functions of the Spearman and Kendall correlation measures. \textit{Statistical Methods and Applications}, 19: 497-515
\bibitem{Hoeffding} Hoeffding, W. (1948). A non-parametric test of independence. \textit{Annals of Mathematical Statistics}, 19(4):546-557
\bibitem{BD} Bergsma, W. \& Dassios, A. (2014). A consistent test of independence based on a sign covariance related to Kendall's tau. \textit{Bernoulli}, 20(2): 1006-1028
\bibitem{BKR} Blum, J. R., Kiefer, J. \& Rosenblatt, M. (1961). Distribution free tests of independence based on the sample distribution function. \textit{Annals of Mathematical Statistics}, 32(2):485-498
\bibitem{chatterjee} Chatterjee, S. (2021). A new coefficient of correlation. \textit{Journal of the American Statistical Association}, 116(536): 2009-2022
\bibitem{Zhang24} Zhang, Q. (2024). On relationships between Chatterjee's and Spearman's correlation coefficients. \textit{Communication in Statistics - Theory \& Methods}. In press
\bibitem{ShiHan21} Shi, H., Drton, M. \& Han, F. (2021). On the power of Chatterjee's rank correlation. \textit{Biometrika}, 109(2): 317-333
\bibitem{LinHan23} Lin, Z. \& Han, F. (2023). On boosting the power of Chatterjee's rank correlation. \textit{Biometrika}, 110(2): 283-299
\bibitem{auddy} Auddy, A., Deb, N. \&  Nandy, S. (2024). Exact detection thresholds and minimax optimality of Chatterjee's correlation coefficient. \textit{Bernoulli}, 30(2): 1640-1668
\bibitem{azadkia.chatterjee} Azadkia, M. \& Chatterjee, S. (2021). A simple measure of conditional dependence. \textit{Annals of Statistics}, 49(6): 3070-3102
\bibitem{LinHan22} Lin, Z. \& Han, F. (2022). Limit theorems of Chatterjee's rank correlation. Available at arXiv:2204.08031
\bibitem{Zhang23} Zhang, Q. (2023). On the asymptotic null distribution of the symmetrized Chatterjee's correlation coefficient. \textit{Statistics \& Probability Letters}, 194
\bibitem{CaoBickel} Cao, S. \& Bickel, P. (2020). Correlations with tailored extremal properties. Available at arXiv:2008.10177
\bibitem{Deb.etal} Deb, N., Ghosal, P. \& Sen, B. (2020). Measuring association on topological spaces using kernels and geometric graphs. Available at arXiv:2010.01768
\bibitem{Huang} Huang, Z., Deb, N. \& Sen, B. (2020). Kernel partial correlation coefficient - a measure of conditional dependence. Available at arXiv:2012.14804v1
\bibitem{ShiHan24} Shi, H., Drton, M. \& Han, F. (2024). On Azadkia-Chatterjee's conditional dependence coefficient. \textit{Bernoulli}, 30(2): 851-877
\bibitem{HanHuang} Han, F. \& Huang, Z. (2022). Azadkia-Chatterjee's correlation coefficient adapts to manifold data. Available at arXiv:2209.11156
\bibitem{chatterjeenet} Chatterjee, S. \& Vidyasagar, M. (2022). Estimating large causal polytree skeletons from small samples. Available at arXiv:2209.07028
\bibitem{chatterjee.survey} Chatterjee, S. (2022). A survey of some recent developments in measures of association. Available at arXiv:2211.04702
\bibitem{dss} Dette, H., Siburg, K.F. \& Stoimenov, P.A. (2013). A copula-based non-parametric measure of regression dependence. \textit{Scandinavian Journal of Statistics}, 40(1): 21-41
\bibitem{Han17} Han, F., Chen, S. \& Liu, H.  (2017). Distribution-free tests of independence in high dimensions. \textit{Biometrika}, 104(4):813-828.
\bibitem{chatterjee.clt} Chatterjee, S. (2008). A new method of normal approximation. \textit{Annals of Probability}, 36(4):1584-1610.
\bibitem{skvector} Grothe, O., Schnieders, J. \& Segers, J.  (2014). Measuring association and dependence between random vectors. \textit{Journal of Multivariate Analysis}, 123(2014): 96-110
\bibitem{KS1} Arndt, S., Turvey, C. \& Andreasen, N.  (1999). Correlating and predicting psychiatric symptom ratings: Spearman's r versus Kendall's tau correlation. \textit{Journal of Psychiatric Research}, 33(2):97-104.
\bibitem{spellman} Spellman, P.T. et al. (1998). Comprehensive identification of cell cycle-regulated genes of the yeast \textit{Saccharomyces cerevisiae} by microarray hybridization. \textit{Molecular Biology of the Cell},  9(12): 3273-3297
\bibitem{mic} Reshef et al. (2011). Detecting novel associations in large datasets. \textit{Science}, 334(6062): 1518-1524
\bibitem{salespeople} Johnson, R. \& Wichern, D. (2002). \textit{Applied Multivariate Statistical Analysis}, Prentice Hall India Learning Private Limited.
\bibitem{dc} Sz\'{e}kely, G. \& Rizzo, M. (2013). The distance correlation t-test of independence in high dimension. \textit{Journal of Multivariate Analysis}, 117: 193-213
\bibitem{dc1} Zhang, Q. (2018). A powerful nonparametric method for detecting differentially co-expressed genes: distance correlation screening and edge-count test. \textit{BMC Systems Biology}, 12(58): 1-16
\bibitem{dc2} Zhang, Q. \& Dao, T. (2020). A distance based multisample test for high-dimensional compositional data with an application to the human microbiome. \textit{BMC Bioinformatics}, 21(205)
\bibitem{dc3} Zhang, Q. (2025). On the properties of distance covariance for categorical data: Robustness, sure screening, and approximate null distributions. \textit{Scandinavian Journal of Statistics}, 52(7): 777-804
\bibitem{angus} Angus, J.E.. (1995). A coupling proof of the asymptotic normality of the permutation oscillation. \textit{Probability in the Engineering and Informational Science}, 9:615-621
\bibitem{Hajek} H\'{a}jek, J.  (1968). Asymptotic normality of simple linear rank statistics under alternatives. \textit{Annals of Mathematical Statistics}, 39(2): 325-346.
\end{thebibliography}
\end{document}